\documentclass[10pt, journal, twoside]{IEEEtran}%a4paper,
\IEEEoverridecommandlockouts
\usepackage{ifpdf}
\usepackage[pdftex]{graphicx}
\usepackage{amsmath, amsfonts, amssymb, bm}
\usepackage[english, onelanguage,ruled,vlined, longend, titlenotnumbered]{algorithm2e}
\usepackage{algorithmic}
\usepackage{array}
\usepackage{url}
\usepackage[usenames, dvipsnames]{xcolor}
\usepackage[utf8]{inputenc}
\usepackage[T1]{fontenc}
\usepackage[compress, nospace]{cite}
\usepackage{siunitx}
\usepackage{pgfplots, tikz}
\usepackage{tkz-berge}
\usetikzlibrary{plotmarks, patterns, arrows, automata, petri, topaths}
\usepackage[english, plain]{fancyref}
\newcommand*{\fancyrefsubseclabelprefix}{subsec}
\frefformat{plain}{\fancyrefsubseclabelprefix}{Subsection~#1}
\newcommand*{\fancyrefalglabelprefix}{alg}
\frefformat{plain}{\fancyrefalglabelprefix}{Algorithm~#1}
\frefformat{plain}{\fancyrefeqlabelprefix}{(#1)}
\usetikzlibrary{shapes.geometric}
\newcommand{\bfA}{\mathbf{A}}
\newcommand{\sfA}{\mathsf{A}}
\newcommand{\tbfA}{\tilde{\mathbf{A}}}

\newcommand{\bfa}{\mathbf{a}}

\newcommand{\bfb}{\mathbf{b}}

\newcommand{\bfB}{\mathbf{B}}
\newcommand{\rmc}{\mathrm{c}}
\newcommand{\bfC}{\mathbf{C}}
\newcommand{\tbfC}{\tilde{\mathbf{C}}}
\newcommand{\bbC}{\mathbb{C}}
\newcommand{\calC}{\mathcal{C}}
\newcommand{\bfd}{\mathbf{d}}
\newcommand{\hbfd}{\hat{\mathbf{d}}}

\newcommand{\bfD}{\mathbf{D}}

\newcommand{\bfE}{\mathbf{E}}
\newcommand{\tbfE}{\tilde{\mathbf{E}}}
\newcommand{\btbfE}{\bar{\tilde{\mathbf{E}}}}
\newcommand{\calE}{\mathcal{E}}

\newcommand{\sfF}{\mathsf{F}}
\newcommand{\bfg}{\mathbf{g}}

\newcommand{\bbfg}{\bar{\mathbf{g}}}
\newcommand{\hbfg}{\hat{\mathbf{g}}}
\newcommand{\bfG}{\mathbf{G}}
\newcommand{\hbfG}{\hat{\mathbf{G}}}
\newcommand{\bbfG}{\bar{\mathbf{G}}}
\newcommand{\calH}{\mathcal{H}}
\newcommand{\sfH}{\mathsf{H}}
\newcommand{\bfI}{\mathbf{I}}
\newcommand{\rmj}{\mathrm{j}}

\newcommand{\bfm}{\mathbf{m}}
\newcommand{\tbfm}{\tilde{\mathbf{m}}}
\newcommand{\hbfm}{\hat{\mathbf{m}}}
\newcommand{\htbfm}{\hat{\tilde{\mathbf{m}}}}
\newcommand{\bfM}{\mathbf{M}}
\newcommand{\tbfM}{\tilde{\mathbf{M}}}
\newcommand{\bfn}{\mathbf{n}}
\newcommand{\rmn}{\mathrm{n}}

\newcommand{\bfN}{\mathbf{N}}
\newcommand{\calN}{\mathcal{N}}
\newcommand{\bfp}{\mathbf{p}}

\newcommand{\hbfr}{\hat{\mathbf{r}}}
\newcommand{\chbfr}{\check{\hat{\mathbf{r}}}}
\newcommand{\bfR}{\mathbf{R}}
\newcommand{\bbR}{\mathbb{R}}
\newcommand{\hbfR}{\hat{\mathbf{R}}}

\newcommand{\bfs}{\mathbf{s}}

\newcommand{\calS}{\mathcal{S}}
\newcommand{\bfT}{\mathbf{T}}
\newcommand{\sfT}{\mathsf{T}}
\newcommand{\bfu}{\mathbf{u}}
\newcommand{\bfU}{\mathbf{U}}

\newcommand{\calU}{\mathcal{U}}

\newcommand{\bfV}{\mathbf{V}}
\newcommand{\cbfv}{\check{\mathbf{v}}}
\newcommand{\cbfV}{\check{\mathbf{V}}}
\newcommand{\bfx}{\mathbf{x}}

\newcommand{\bfW}{\mathbf{W}}

\newcommand{\bfy}{\mathbf{y}}
\newcommand{\bbfy}{\bar{\mathbf{y}}}

\newcommand{\bfz}{\mathbf{z}}

\newcommand{\boalpha}{\boldsymbol{\alpha}}
\newcommand{\hboalpha}{\hat{\boldsymbol{\alpha}}}

\newcommand{\boGamma}{\boldsymbol{\Gamma}}

\newcommand{\boomega}{\boldsymbol{\omega}}
\newcommand{\boOmega}{\boldsymbol{\Omega}}

\newcommand{\boPsi}{\boldsymbol{\Psi}}

\newcommand{\bosigma}{\boldsymbol{\sigma}}
\newcommand{\boSigma}{\boldsymbol{\Sigma}}
\newcommand{\cbosigma}{\check{\boldsymbol{\sigma}}}

\newcommand{\hbosigma}{\hat{\boldsymbol{\sigma}}}

\newcommand{\botheta}{\boldsymbol{\theta}}

\newcommand{\boxi}{\boldsymbol{\xi}}
\newcommand{\boXi}{\boldsymbol{\Xi}}

\newcommand{\bfRzero}{\mathbf{R}^{\TRUE}}

\newcommand{\hbfRzero}{\hat{\mathbf{R}}^{\TRUE}}

\newcommand{\boSigman}{\boldsymbol{\Sigma}^{\mathrm{n}}}

\DeclareMathOperator{\diag}{diag}
\DeclareMathOperator{\sign}{sign}
\DeclareMathOperator{\vect}{vec}
\DeclareMathOperator{\vectdiag}{vecdiag}
\DeclareMathOperator*{\argmin}{arg\,min}

\newcommand{\bfun}{\mathbf{1}}
\newcommand{\bfzero}{\mathbf{0}}

\newcommand{\OUT}{\text{\tiny U}}
\newcommand{\TRUE}{\text{\tiny K}}

\RequirePackage[l2tabu, orthodox]{nag}
\begin{document}
\title{Parallel Calibration for Sensor Array \\ Radio Interferometers}

\author{\IEEEauthorblockN{Martin Brossard, Mohammed Nabil El Korso, Marius Pesavento, \IEEEmembership{Member, IEEE,} \\ Rémy Boyer, \IEEEmembership{Member, IEEE}, Pascal Larzabal, \IEEEmembership{Member, IEEE} and Stefan J. Wijnholds, \IEEEmembership{Senior Member, IEEE}}

\thanks{Martin Brossard and Pascal Larzabal are with SATIE, UMR 8029, École Normale Supérieure de Cachan, Cachan, France (e-mail: martin.brossard@ens-cachan.fr, pascal.larzabal@satie.ens-cachan.fr).}
\thanks{Martin Brossard and Marius Pesavento are with Communication Systems Group, Technische Universität, Darmstadt, Germany (e-mail: mpesa@nt.tu-darmstadt.de).}
\thanks{Mohammed Nabil El Korso is with University of Paris Ouest Nanterre La Défense, IUT de Ville d’Avray, LEME EA 4416, France (e-mail: m.elkorso@u-paris10.fr).}
\thanks{Rémy Boyer is with University of Paris-Sud, Laboratoire des Signaux et Systèmes (L2S), Gif-Sur-Yvette, France (e-mail: remy.boyer@l2s.centralesupelec.fr).} 
\thanks{Stefan J. Wijnholds is with the Netherlands Institute for Radio Astronomy (ASTRON), P.O. Box 2, NL-7990 AA, Dwingeloo, The Netherlands (e-mail: wijnholds@astron.nl).}
\thanks{This work was supported by MAGELLAN(ANR-14-CE23-0004-01) and by the iCODE institute, research project of the IDEX Paris-Saclay. This work is also funded by IBM, ASTRON, the Dutch Ministry of Economic Affairs and the Province of Drenthe.}
}
\maketitle

\begin{abstract}
In order to meet the theoretically achievable imaging performance, calibration of modern radio interferometers is a mandatory challenge, especially at low frequencies. In this perspective, we propose a novel parallel iterative multi-wavelength calibration algorithm. The proposed algorithm estimates the apparent directions of the calibration sources, the directional and undirectional complex gains of the array elements and their noise powers, with a reasonable computational complexity. Furthermore, the algorithm takes into account the specific variation of the aforementioned parameter values across wavelength. Realistic numerical simulations reveal that the proposed scheme outperforms the mono-wavelength calibration scheme and approaches the derived constrained Cramér-Rao bound even with the presence of non-calibration sources at unknown directions, in a computationally efficient manner.
\end{abstract}


\section{Introduction}
\IEEEPARstart{A}{dvanced} radio interferometers, as the existing LOw Frequency ARray (LOFAR) \cite{Haa2013} and the future Square Kilometre Array (SKA) \cite{Dew2009}, form large sensor arrays, which are constituted of many small antenna elements. As an example, the LOFAR consists of 50 stations, mainly located across the Netherlands. Each station is a closed packed sensor array, composed of at least 96 low-band antennas (30-\SI{90}{MHz}) and 48 high-band antennas (110-\SI{240}{MHz}). Such interferometers offer a large aperture size and deliver large amounts of data in order to reach high performance in terms of resolution, sensitivity and survey speed \cite{Dew2009}. Nevertheless, to achieve the theoretical optimal performance bounds, a plethora of signal processing challenges must be treated \cite{Wij2010, Wij2014}. This covers calibration, image synthesis and data reduction. In this paper, we focus on calibration issues by designing a computationally efficient parallel algorithm. Calibration procedures devised for such radio interferometers must estimate: i) the gain response and noise power of each antenna \cite{Wij2009, Wijthe, Ven2013, Van2007_2}; and ii) the propagation disturbances, especially the phase delays caused by the ionosphere, which scale with wavelength \cite{Tho2001, Vanthe}.

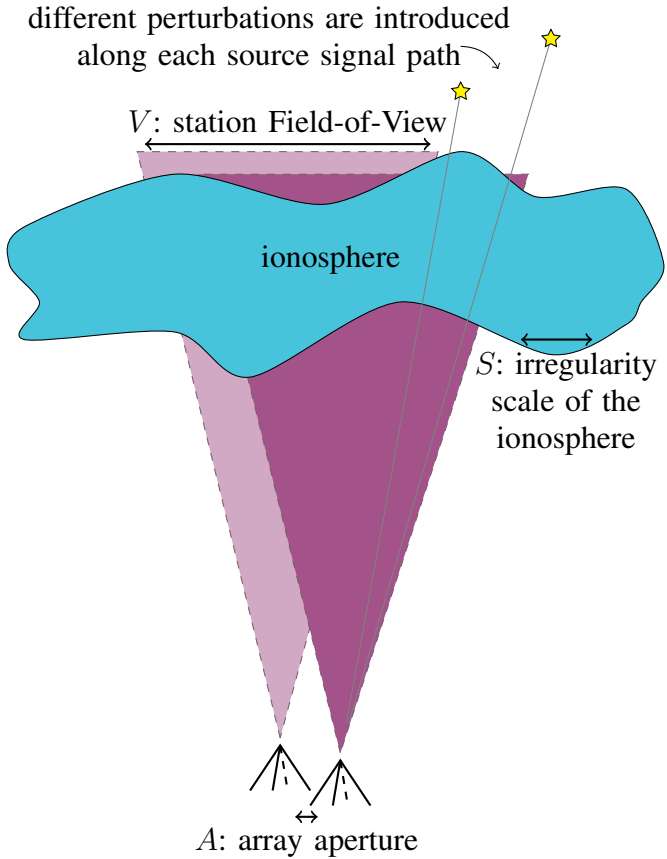
\begin{figure}
 \centering
 \begin{tikzpicture}
        \coordinate (f1) at (0.4,0.7);
        \coordinate (f2) at (-1.5,8.5);
        \coordinate (f3) at (2.5,8.5);
        \coordinate (f4) at (1.2,0.5);
        \coordinate (f5) at (-0.7,8.2);
        \coordinate (f6) at (3.7, 8.2);
        \coordinate (i1) at (-3, 6);
        \coordinate (i2) at (-2.8, 6.5);
        \coordinate (i3) at (-3.2, 7);
        \coordinate (i4) at (-3, 7.5); %angle1
        \coordinate (i5) at (-1, 8.2);
        \coordinate (i51) at (1, 7.8);
        \coordinate (i6) at (2.8, 8.5);
        \coordinate (i7) at (3.8, 7.9);
        \coordinate (i8) at (5, 8); %angle2
        \coordinate (i9) at (5.5, 7);
        \coordinate (i10) at (5.2, 6.5);
        \coordinate (i11) at (5, 6.2);
        \coordinate (i111) at (4, 5.8);
        \coordinate (i12) at (2, 6.5);
        \coordinate (i121) at (0, 5.5);
        \coordinate (i13) at (-1, 6.1);
        \coordinate (t4) at (2.5, 3);
        \coordinate (t3) at (4.2, 5.15);
        \coordinate (e1) at (4.0, 10.0);
        \coordinate (e2) at (2.8, 9.3);
        \large
            \draw[color=black,thick] (0,0) -- +(0.4,0.6) -- +(0.8, 0); %antenne 1
            \draw[color=black,thick] (0.4,0.6) -- +(-0.1,-0.6);
            \draw[color=black,thick,dashed] (0.4,0.6) -- +(+0.1,-0.6);
            \draw[color=black,thick] (0.8,-0.2) -- +(0.4,0.6) -- +(0.8, 0); %antenne 2 
            \draw[color=black,thick] (1.2,0.4) -- +(-0.1,-0.6);
            \draw[color=black,thick,dashed] (1.2,0.4) -- +(+0.1,-0.6);
            \fill[color=DarkOrchid, opacity=0.5] (f1) -- (f2) -- (f3) -- cycle;%FoV 1
            \draw[color=DarkOrchid!50!black,dashed] (f1) -- (f2) -- (f3) -- cycle;
            \fill[color=DarkOrchid] (f4) -- (f5) -- (f6) -- cycle;%FoV 2
            \draw[color=DarkOrchid!50!black,dashed] (f4) -- (f5) -- (f6) -- cycle;
            \fill[color=SkyBlue] plot [smooth cycle] coordinates {(i1) (i2) (i3) (i4) (i5) (i51) (i6) (i7) (i8) (i9) (i10) (i11) (i111) (i12) (i121) (i13)};%Ionosphere
            \draw[color=black] plot [smooth cycle] coordinates {(i1) (i2) (i3) (i4) (i5) (i51) (i6) (i7) (i8) (i9) (i10) (i11) (i111) (i12) (i121) (i13)};
            
            \draw[color=black,thick, <->] (0.6, -0.35) -- (0.9,-0.35) node[midway, below] {$A$: array aperture}; %texte
            \draw[color=black,thick, <->] (-1.4,8.6) -- (2.4, 8.6) node[midway, above] {$V$: station Field-of-View};
            \draw[color=black,thick, <->] (3.6,6) -- (4.55,6); 
            \draw (0,7.4) node[below right] {ionosphere} node{};
            \draw (t3) node[text width=2.8cm, text centered] {$S$: irregularity scale of the ionosphere} node{};
            \draw[color=gray] (1.31,0.9) -- (e1);
            \draw[color=gray] (1.28,0.9) -- (e2);
            \draw (3.7, 10) node[left, text width=6.6cm, text centered] {different perturbations are introduced along each source signal path};

            \path[color=black, <-,bend right] (3.3,9.6) edge (2.8, 9.9); 
            \draw (e1) node[star, star points=5, star point ratio=1.95, draw,inner sep=1.3pt, fill=yellow]{ };
            \draw (e2) node[star, star points=5, star point ratio=1.95, draw,inner sep=1.3pt, fill=yellow]{ };
            
        \end{tikzpicture}
 \caption{The so-called regime 3, which is considered in this paper, assumes that $V\gg S$ and $A\ll S$. This leads to ionospheric perturbations which are direction dependent (after \cite{Lon2004, Van2007_2}).\label{fig:scenario}}
\end{figure}

Specifically, in this paper, we focus on the regime where all lines of sight toward a source in the sky cross the same ionospheric layer and where the thickness of the ionosphere can be direction dependent \cite{Lon2004}, which is represented in \Fref{fig:scenario} and well adapted for the calibration of a LOFAR station and the future SKA stations as well as the core of these arrays. Consequently, in this regime, the ionospheric phase delays modify the geometric delays and introduce angular-shifts for the source directions \cite{Cot2004, Ven2013}, which are direction and wavelength dependent \cite{Tho2001, Coh2009}. By estimating calibrator shifts (i.e., the difference between the true calibrator directions, known from tables \cite{Ben1962, Baa1977, Kim2008, Bry2009}, and their estimated apparent directions), interpolation methods can be efficiently applied in order to obtain a phase screen model, that captures the  ionospheric delays over the entire Field-of-View \cite{Cot2004}. We emphasize that in addition to the phase screen reconstruction step, the calibration usually involves the estimation of the complex undirectional gains of the antennas, their directional gains toward each calibrator and their noise powers \cite{Wijthe}, for the whole available range of wavelength range, i.e., processing bandwidth.

The characteristics of the calibration sources, i.e., their true/nominal directions and their powers without the effects of the ionosphere nor antenna imperfections, are a priori knowledge which is required to solve such calibration problems \cite{Ven2013}. Based on this knowledge, state-of-the-art calibration algorithms operate mostly in an iterative manner in a mono-wavelength scenario \cite{Wijthe, Ven2013, Wij2009, Sal2014, Sar2014, Kaz2015}. For instance, the (Weighted) Alternating Least Squares approach has been adapted for LOFAR station calibration \cite{Wijthe, Wij2009}, in which closed-form expressions have been obtained for antenna gain and sensor noise power parameters. Nevertheless, such algorithms present three major limitations: i) suboptimality due to the consideration of only one wavelength bin; ii) assumption of centralized processor, i.e., a single compute agent simultaneous accesses all data; and iii) inefficiency regarding to the Direction-of-Arrival (DoA) estimation in the severe radio astronomical contexts.  

Concerning limitation i), most existing calibration schemes \cite{Wijthe, Ven2013, Wij2009, Sal2014, Sar2014, Kaz2015} were designed for calibration of a single wavelength at a time. Smoothness across wavelength is usually enforced post facto by fitting functions to the calibration solutions obtained or by filtering them \cite{Tas2014}. Such approaches may not be optimal, since they do not take advantage of the possibility of cost function optimization over the entire frequency range. To the best of our knowledge, the only recent approach to consider multi-wavelength calibration in the context of large interferometer arrays is the procedure presented in \cite{Yat2015}, which aims to enforce the smoothness of the solutions with the Jones matrix formulation. The procedure presented in \cite{Yat2015} is based on an algorithmic model while we propose to use a physical model.

Furthermore, regarding the limitation ii), the aforementioned state-of-the-art methods typically operate in a centralized hardware architecture, whereas, taking the LOFAR as example, storing and reading all 512 sub-wavelength bands at a single location is challenging. As a solution, distributed and consensus algorithms, mostly based on the Alternating Direction of Multiple Multipliers (ADMM) \cite{Boy2011}, have recently been massively investigated in parametric estimation frameworks \cite{Son2015, Shi2O12, Iut2014, Gol2014, Cha2015, Ers2012, Shi2015_2, Mot2013}. These distributed schemes can operate in various network topologies. We will consider a group of compute agents, where each agent accesses data across a small bandwidth and can only communicates with a fusion center through low data rate channels, as employed in \cite{Yat2015}. This architecture models correctly the situation for radio interferometers, where data for the full observing bandwidth is typically divided into channels and channels are grouped into subbands. 

Finally, regarding the limitation iii), classical subspace methods, such as MUSIC \cite{Sch1986}, have been commonly applied in radio astronomical calibration \cite{Wijthe}. However, these techniques are inefficient in low Signal-to-Noise-Ratio (SNR) scenarios and require knowledge of the exact number of sources in the scene. As an alternative, recent approaches, based on sparse reconstruction methods, came into focus of DoA estimation for fully calibrated arrays \cite{Mal2005, Wei2012, Nor2013} as well as for partially calibrated arrays \cite{Ste2014}. These approaches exhibit the super-resolution property, robustness and computational efficiency, without the aforementioned limitations of subspace-based methods \cite{Mal2005}. However, most methods based on the compressive sensing framework operate in a centralized architecture and are applied in the signal time domain \cite{Bil2014, Kaz2015}. This becomes computationally unfeasible with huge numbers of observations, making such methods unsuitable for radio interferometers context, for which we commonly access only the sample covariance matrix rather than the time signal itself \cite{Ven2013}.

In summary, we propose an iterative algorithm, namely the Parallel Calibration Algorithm (PCA), that focuses on the calibration of a sensor array based radio interferometer, involving its individual antennas and propagation disturbances. In addition, we assume that the sensor array has an arbitrary geometry, identical elements and is simultaneously excited by inaccurately known calibration sources and unknown non-calibration sources. We consider these non-calibration sources as outliers, i.e., as an additional noise term (a.k.a. outliers in our calibration procedure). The proposed PCA overcomes the aforementioned limitations, by: i) reformulating the parametric model in the multi-wavelength scenario in order to exploit wavelength diversity; ii) relying on distributed and consensus algorithms; and iii) adapting the sparse reconstruction methods to the calibration of radio interferometers. From the parallel calibration perspective, the PCA successively estimates the undirectional antenna gains along with the directional and noise parameters for multiple subbands, where we enforce the coherence over the wavelength of the estimates based on physical and astronomical phenomena \cite{Bre2012, Tho2001, Coh2009, Van2007_2}. Furthermore, the sensor noise power estimation considers the presence of non-calibration sources.

The rest of the paper is organized as follows: in \Fref{sec:data}, we formulate the data model and its associated parallel multi-wavelength calibration problem. In \Fref{sec:proposed_algo}, we present the overview of the proposed scheme and then describe its two main alternating steps. The constrained Cramér-Rao bound of the data model is derived in \Fref{sec:cr_bound}. Numerical simulations, in \Fref{sec:simulations}, show the feasibility and superiority of the proposed scheme compared to mono-wavelength calibration. Finally, we give our conclusions in \Fref{sec:conclusion}.

In the following, $\bar{(.)}, (.)^{\sfT}, (.)^{\sfH}, (.)^{\dagger}, (.)^{\odot \alpha}, \Re(.), \Im(.)$ and $[.]_{n}$ denote, respectively, conjugation, transposition, Hermitian transposition, pseudo-inverse, element-wise raising to $\alpha$, real part, imaginary part and the $n$-th element of a vector. The expectation operator is $\calE\{.\}, \circ$ denotes the Khatri-Rao product, $\exp(.)$ and $\odot$ represent the element-wise exponential function and multiplication (Hadamard product), respectively. The operator $\diag(.)$ converts a vector to a diagonal matrix with the vector aligned on the main diagonal, whereas $\vectdiag(.)$ produces a vector from the main diagonal of its entry and $\vect(.)$ converts a matrix to a vector by stacking the columns of its entry. The operators $\left\| . \right\|_{0}, \left\|.\right\|_2$ and $\left\|.\right\|_{\sfF}$ refer to the $l_0$ norm, i.e., the number of non-zero elements of its entry, the $l_2$ and Frobenius norms, respectively. Finally, $\bfx \succeq \bfzero$ means that each element in $\bfx$ is non-negative.

\section{Data Model \& Problem Statement}
\label{sec:data}
\subsection{Covariance Matrix Model}
\label{subsec:covariance}
Consider an array comprised of $P$ elements, with known locations, each referred by its Cartesian coordinates $\boxi_p = \left[x_p, y_p, z_p \right]^{\sfT}$ for $p = \allowbreak 1,\ldots,P$, that we stack in $\boXi = \left[\boxi_1,\ldots,\boxi_P \right]^{\sfT} \in \bbR^{P \times 3}$. This array is exposed to $Q$ known strong calibration sources and  $Q^{\OUT}$ unknown weak non-calibration sources. Let $\bfD^{\TRUE} = \left[ \bfd_{1}^{\TRUE} , \ldots , \bfd_{Q}^{\TRUE} \right]  \in \bbR^{3 \times Q}$ and $\bfD^{\OUT} = \big[ \bfd^{\OUT}_{1} , \ldots , \bfd_{Q^{\OUT}}^{\OUT} \big] \in \bbR^{3 \times Q^{\OUT}}$ denote the known (\emph{true/nominal}) calibrator  direction cosines and unknown non-calibrator  direction cosines, respectively, in which each source direction  $\bfd = \left[d_{l},d_{m},d_{n}\right]^{\sfT}$ can be uniquely described by a couple $(d_{l},d_{m})$, since $d_{n} = \sqrt{1-d_{l}^2-d_{m}^2}$ \cite{Tho2001, Wijthe}. The ionosphere introduces an unknown angular-shift for each source direction \cite{Cot2004, Wij2010, Coh2009}, depending on the wavelength $\lambda$, which is related to the frequency $f = \frac{\rmc}{\lambda}$, with $\rmc$ denoting the light speed. Consequently, we distinguish between the  unknown \emph{apparent} directions w.r.t. the calibrators, denoted by $\bfD_{\lambda} = \left[ \bfd_{\lambda, 1} ,\ldots , \bfd_{\lambda, Q} \right]$, and their \emph{true/nominal} known directions $\bfD^{\TRUE}$, i.e., without the propagation disturbances. 

In the following, we describe the signal for one wavelength bin. Under the narrowband assumption, the steering vector $\bfa_{\lambda}(\bfd)$  toward the direction $\bfd$ at wavelength $\lambda$ is given by
\begin{equation}
\bfa_{\lambda}(\bfd) = \bfa_{\lambda}(l, m) =  \frac{1}{\sqrt{P}} \exp \left( - \rmj \frac{2\pi}{\lambda} \boXi \bfd \right) \text{,}
\end{equation}
that we gather for multiple directions in the steering matrix
\begin{equation}
 \bfA_{\bfD_{\lambda}} = \frac{1}{\sqrt{P}} \exp \left( -\rmj \frac{2\pi}{\lambda} \boXi \bfD_{\lambda} \right) \text{.}
\end{equation}

As in \cite{Wijthe}, we assume that all antennas have identical directional responses. Their directional gain responses (and propagation losses) are modeled by two diagonal matrices, $\boGamma_{\lambda} \in \allowbreak \bbC^{Q \times Q}$ and $\boGamma^{\OUT}_{\lambda} \in \allowbreak \bbC^{Q^{\OUT} \times Q^{\OUT}}$, toward the calibration and non-calibration sources, respectively.

The received signals from each antenna are divided into narrow subbands and stacked, leading to the vector
\begin{equation}
 \bfx_{\lambda}(n) = \bfG_{\lambda} \big[ \bfA_{\bfD_{\lambda}} \boGamma_{\lambda} \bfs_{\lambda}(n)    + \bfA_{\bfD^{\OUT}_{\lambda}} \boGamma^{\OUT}_{\lambda} \bfs^{\OUT}_{\lambda}(n) \big] +  \bfn_{\lambda}(n) \label{eq:signal}\text{,}
\end{equation}
for the $n$-th observation and wavelength $\lambda$, with $[\bfx_{\lambda}(n)]_{p}$ the signal corresponding to the $p$-th antenna, where $\bfG_{\lambda} = \diag(\bfg_{\lambda}) \allowbreak \in \bbC^{P \times P}$ models the undirectional antenna gains, with $[\bfg_{\lambda}]_{p}$ the undirectional antenna gain for the $p$-th antenna, $\bfs_\lambda(n) \in \allowbreak \bbC^{Q}$ and  $\bfs^{\OUT}_{\lambda} (n) \in \allowbreak \bbC^{Q^{\OUT}}$ represent, respectively, the i.i.d. calibrator and non-calibrator signals, with $[\bfs_\lambda(n)]_{q}$ and $[\bfs^{\OUT}_{\lambda} (n)]_{q'}$, respectively, the signal corresponding to the $q$-th calibrator and $q'$-th non-calibrator, whereas $\bfn_{\lambda}(n) \sim \mathcal{CN}(\bfzero, \boSigman_{\lambda})$ denotes the i.i.d. noise vector, with $[\bfn_{\lambda}(n)]_{p}$ the thermal noise for the $p$-th antenna \cite{Ven2013}. Let $\boSigma_{\lambda}  = \diag\left(\bosigma_{\lambda}\right) \in \allowbreak \bbR^{Q \times Q}, \boSigma^{\OUT}_{\lambda}  = \diag\left(\bosigma^{\OUT}_{\lambda}\right) \in \allowbreak \bbR^{Q^{\OUT} \times Q^{\OUT}}$ and $\boSigman_{\lambda} = \diag\left(\bosigma^{\rmn}_{\lambda}\right) \allowbreak \in \bbR^{P \times P}$ be the diagonal covariance matrices for the calibrators, non-calibration sources and sensor noises, respectively, and assume that the sources are statistically independent from each other. Consequently, $\bfs_{\lambda}(n)\sim \mathcal{CN}(\bfzero, \boSigma_{\lambda}), \bfs^{\OUT}_{\lambda}(n) \sim \mathcal{CN}(\bfzero, \boSigma_{\lambda}^{\OUT})$ and the covariance matrix $\bfR_{\lambda} = \allowbreak \calE \left\lbrace \bfx_{\lambda} \bfx_{\lambda}^{\sfH} \right\rbrace  $ of the observations corresponding to model \fref{eq:signal} is given by 
\begin{equation}
\begin{aligned}
  \bfR_{\lambda} &= \bfE_{\bfD_{\lambda}} \bfM_{\lambda} \bfE_{\bfD_{\lambda}}^{\sfH} + \bfR^{\OUT}_{\lambda} + \boSigman_{\lambda} \text{,} \label{eq:cov}
\end{aligned}
\end{equation}
in which
\begin{align}
 \bfE_{\bfD_{\lambda}} &= \bfG_{\lambda} \bfA_{\bfD_{\lambda}} \boSigma^{\frac{1}{2}}_{\lambda} \text{,}\\
 \bfM_{\lambda} &= \boGamma_{\lambda} \boGamma_{\lambda}^{\sfH} =  \diag\left(\bfm_{\lambda}\right) \text{,}
\end{align}
and where we have defined the unknown covariance matrix for the non-calibration sources as 
\begin{equation}
 \bfR^{\OUT}_{\lambda} = \bfG_{\lambda}\bfA_{\bfD^{\OUT}_{\lambda}}\boGamma^{\OUT}_{\lambda}\boSigma^{\OUT}_{\lambda} \big(\bfG_{\lambda}\bfA_{\bfD^{\OUT}_{\lambda}}\boGamma^{\OUT}_{\lambda} \big)^{\sfH}\text{.}
\end{equation}

In order to overcome the scaling ambiguities in the observation model \fref{eq:cov} \cite{Van2007_2}, we consider the following commonly used assumptions in radio astronomy \cite{Wijthe, Ven2013}: i) to resolve the phase ambiguity of $\bfg_{\lambda}$, we take its first element as the phase reference; ii) $\bfm_{\lambda}$ shares a common scalar factor with $\bfg_{\lambda}$ and consequently, we assume that the directional gain towards the first calibration source is known/fixed; and iii) when solving for the calibrator directions, a common rotation of all steering vectors can be compensated by the undirectional gain phase solution. We therefore fix the direction of the first calibration source at its known position.

\subsection{Model Effects of the Wavelength on Antenna Gains, Source Direction Shifts and Source Powers}
\label{subsec:wavelength}
In the radio astronomy context, the antenna and source parameters of the covariance matrix are commonly assumed wavelength dependent \cite{Ven2013, Van2007_2}. Consequently, we assume smooth or/and known variations of the parameters $\bfg_{\lambda}, \boGamma_{\lambda}, \boSigma_{\lambda}, \boSigma^{\OUT}_{\lambda}$ and $\boSigman_{\lambda}$ in \fref{eq:cov} over $\lambda$, as commonly used in recent astronomy applications \cite{Tas2014, Yat2015}. We summarize the particular behavior of the underlying parameters as follows:
\begin{itemize}
 \item The undirectional gains, $\bfg_{\lambda}$, vary smoothly over $\lambda$. Common models for characterizing these behaviors consist of classical polynomials of power law over $\lambda$ \cite{Yat2015, Van2007_2}.
  \item The directional gains, $\boGamma_{\lambda}$, are inversely proportional to $\lambda$, i.e., $\boGamma_{\lambda} \propto \lambda^{-1}$, as observed in practice \cite{Bre2012}. Note that the proposed algorithm can be straightforwardly adapted with another given behavior (including the extreme case of a constant behavior across the wavelength range).
 \item As a consequence of the ionospheric delays, that are at the origin of the directional shifts, the shifts are proportional to $\lambda^2$ \cite{Tho2001, Coh2009, Bre2012}.
 \item The source powers, $\boSigma_{\lambda}$ and $\boSigma^{\OUT}_{\lambda}$, vary commonly with a power law with different spectral indexes.  We consider the calibrator powers, $\boSigma_{\lambda}$, to be known from tables, e.g., \cite{Ben1962, Baa1977, Kim2008, Bry2009}.
 \item The antenna noise, $\boSigman_{\lambda}$, does not follow a smooth behavior w.r.t. $\lambda$ and is assumed i.i.d. over wavelength. Nevertheless, if particular coherence models for the noise covariances is available, this knowledge can be incorporated in the proposed algorithm in a straightforward manner.
\end{itemize}

\subsection{Joint Parameter Estimation Problem}
\label{subsec:joint}
In this subsection, we formulate the calibration problem as the estimation of the parameter vector of interest, $\bfp$, defined as
\begin{equation}
\bfp = \left[\bfp_{\lambda_1}^{\sfT}, \ldots, \bfp_{\lambda_J}^{\sfT} \right]^{\sfT} \text{,}
\end{equation}
in which $\bfp_{\lambda} = \big[\bfg_{\lambda}^{\sfT}, \bfd^{\sfT}_{\lambda, 1}, \ldots, \bfd^{\sfT}_{\lambda, Q}, \bfm_{\lambda}^{\sfT}, \bosigma^{\rmn \sfT}_{\lambda} \big]^{\sfT}$, from $J$ sample covariance matrices
\begin{equation}
 \bigg\lbrace \hbfR_{\lambda} = \frac{1}{N} \sum_{n=1}^N \bfx_{\lambda}(n) \bfx_{\lambda}^{\sfH}(n)\bigg\rbrace_{\lambda \in \Lambda}\text{,} \label{eq:sample_cov}
\end{equation}
where $\Lambda = \left\lbrace \lambda_1, \ldots, \lambda_{J} \right\rbrace$ represents the set of the $J$ available wavelengths for the whole network.

\begin{figure}
 \centering
 \begin{tikzpicture}[scale=0.75,transform shape]
\large
 \draw[dotted, thick] (3,0) -- (5,0);
 \draw[dotted, thick] (3,-0.8) -- (5,-0.8);
\GraphInit[vstyle=Normal]
\renewcommand*{\VertexLineColor}{white}
  \Vertex[x=0,y=-0.8, L={$\lbrace\hbfR_{\lambda}\rbrace_{\lambda \in \Lambda_1}$}]{R1}
  \Vertex[x=8,y=-0.8, L={$\lbrace \hbfR_{\lambda} \rbrace_{\lambda \in \Lambda_Z}$}]{RZ}
  \renewcommand*{\VertexLineColor}{black}
  \Vertex[x=0,y=0, L=$\sfA_1$]{A1}
  \Vertex[x=8,y=0, L=$\sfA_Z$]{AZ}

  \Vertex[x=4,y=5,L={Fusion  center}]{FC}
   \tikzstyle{EdgeStyle}=[post]
   \Edge[label={$\left\lbrace \bfg_{\lambda}, \tbfm_{\lambda}\right\rbrace_{\lambda \in \Lambda_1}$}, style={bend left}](A1)(FC)
   \Edge[label={$\boalpha, \bfm, \bfD_{\lambda_{0}}$}, style={bend left}](FC)(A1)
   \Edge[label={$\left\lbrace \bfg_{\lambda}, \tbfm_{\lambda}\right\rbrace_{\lambda \in \Lambda_Z}$}, style={bend right}](AZ)(FC)
   \Edge[label={$\boalpha, \bfm, \bfD_{\lambda_{0}}$}, style={bend right}](FC)(AZ)
\end{tikzpicture}
 \caption{Parallel calibration uses $Z$ agents, each operates on data provided from a subset of available data. Information is exchanged between the agents via a fusion center. The total amount of information transferred is considerably lower than the amount of available data.\label{fig:dessin}}
\end{figure}
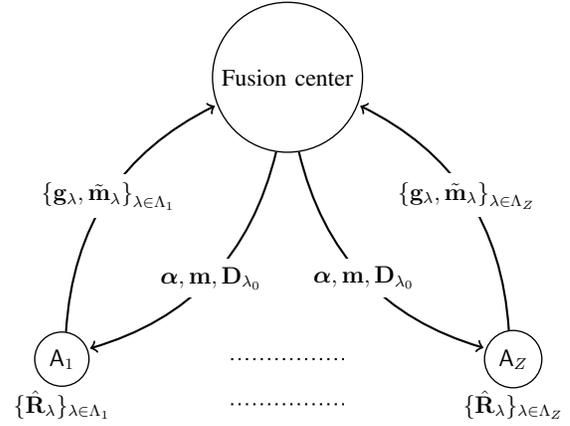

Data parallelism across wavelength is inherent in radio astronomical observations, which are recorded as multiple channels at different wavelengths \cite{Yat2015}. Thus, we consider that data is not centralized but distributed across a network. This network consists of: i) one fusion center, that does not access data; and ii) $Z$ compute agents. The $z$-th agent, $\sfA_z$, can only access data for a subset $\Lambda_{z} \subset \Lambda$ of $J_z<J$ subbands, and for each available wavelength, its associated sample covariance matrix is accessible for exactly one agent. Moreover, the agents cannot transfer information between themselves, but can only communicate with the fusion center at a low communication rate, as shown in \Fref{fig:dessin}.
 
Note that the estimation of the unknown matrices $\bfR^{\OUT}_{\lambda}$ represents the imaging step which is beyond the scope of the paper \cite{Wijthe, Ven2013, Van2007, Tho2001}. Image synthesis \cite{Tay1998, Wia2009, Car2014, Fer2014, Gar2015} is usually performed as a separate step after the calibration and can be complemented by the proposed calibration approach. The main reason for this two-step procedure is that the calibration step is usually carried out based on a point source model assumption (unlike the imaging step) with a known number of strong calibrators, whereas, the effect of an unknown number of the weakest (non-calibration) sources can be assumed absorbed by the noise component.

\section{Proposed Parallel Calibration Algorithm}
\label{sec:proposed_algo}
\subsection{Overview of the Proposed Parallel Calibration Algorithm}
\label{subsec:overview}

\begin{algorithm}[b]
 \KwIn{$\big\lbrace \hbfR_{\lambda} \big\rbrace_{\lambda \in \Lambda}, \eta_{\bfp}$\;}
 \textbf{Init:} set $i = 0, \bfg_{\lambda} = \bfg_{\lambda}^{[0]}$, $ \bfD_{\lambda}=\bfD^{\TRUE}, \bfm_{\lambda} = \bfm_{\lambda}^{[0]}, \boOmega_{\lambda} = \bfun_{P \times P}$\;
  \While{$\left\| \bfp^{[i-1]}-\bfp^{[i]} \right\|_2 \ge \left\| \bfp^{[i]} \right\|_2 \eta_{\bfp}$}{
  \nl $i = i+1$\;
  \nl Estimate $\bfg_{\lambda}^{[i]}$ \label{lst:line:g} with \fref{alg:gain_estimation}\;
  \nl Estimate $\bfD^{[i]}_{\lambda}, \bfm_{\lambda}^{[i]}, \bosigma^{\rmn[i]}_{\lambda}$ \label{lst:line:doa} with \fref{alg:iht}\;
  \nl Update locally $\boOmega_{\lambda}^{[i]} = \left( \bosigma^{\rmn[i]}_{\lambda} \bosigma_{\lambda}^{\rmn[i] \sfT} \right)^{\odot -\frac{1}{2}}$\;}
 \KwOut{$\hat{\bfp} =  \big[\bfp_{\lambda_1}^{[i]\sfT}, \ldots, \bfp_{\lambda_J}^{[i]\sfT} \big]^{\sfT}$\;}
 \caption[]{Parallel Calibration Algorithm\label{alg:proposed}}
\end{algorithm}

It is well established that a statistically efficient estimator can be obtained via the Maximum Likehood formulation. However, from a computational viewpoint, its exact evaluation appears to be intractable in the radio astronomy context \cite{Wijthe}. With a large number of samples, statistically efficient estimators can be devised using the Weighting Least Squares approach. In this context, we define the local cost function to minimize, for each $\lambda \in \Lambda$, as: $\kappa_{\lambda}(\bfp_{\lambda}) = \left\| \bfW_{\lambda}^{-\frac{1}{2}} \left( \bfR_{\lambda}(\bfp_{\lambda}) - \hbfR_{\lambda} \right) \bfW_{\lambda}^{-\frac{1}{2}} \right\|_{\sfF}^2$ \cite{Wij2009}, with 
\begin{equation}
 \bfR_{\lambda}(\bfp_{\lambda}) = \bfE_{\bfD_{\lambda}}\bfM_{\lambda}\bfE^{\sfH}_{\bfD_{\lambda}} + \boSigman_{\lambda}
\end{equation}
denoting the covariance matrix when the contribution of the non-calibrators is included in the noise, and $\bfW_{\lambda}$ being the weighting matrix. The optimal weighting matrix for Gaussian noise is the inverse of the covariance of the residuals \cite{Ott1998}, which is generally unknown. In radio astronomy, sources (including the calibration sources), are typically much weaker than the antenna noise \cite{Sal2014}, so the covariance matrix can be approximated by $\bfR_{\lambda} \approx \boSigma^{\rmn}_{\lambda}$. Since the array consists of identical elements and mutual coupling can usually be ignored, it is commonly assumed that $\boSigma^{\rmn}_{\lambda} = \allowbreak \diag \left(\bosigma^{\rmn}_{\lambda} \right) \approx  \allowbreak \sigma_{\lambda}^{\rmn} \bfI$. Consequently, we consider in our alternating algorithm $\bfW_{\lambda} = \bfI$ as an initialization and refine it with $\bfW_{\lambda} = \boSigma^{\rmn}_{\lambda}$ once we obtain an estimate of $\boSigma^{\rmn}_{\lambda}$. Since $\boSigma^{\rmn}_{\lambda}$ is diagonal, we rewrite the local cost function, i.e., the cost function associated with the wavelength $\lambda$, as
\begin{equation}
\kappa_{\lambda}(\bfp_{\lambda}) = \left\| \left( \bfR_{\lambda}(\bfp_{\lambda}) - \hbfR_{\lambda} \right) \odot \boOmega_{\lambda} \right\|_{\sfF}^2  \label{eq:kappa:2} \text{,}
\end{equation}
with $\boOmega_{\lambda} = \left( \bosigma^{\rmn}_{\lambda} \bosigma^{\rmn \sfT}_{\lambda} \right)^{\odot -\frac{1}{2}}$. Finally, we define the global cost function as
\begin{equation}
\kappa(\bfp) = \sum_{\lambda \in \Lambda} \kappa_{\lambda}(\bfp_{\lambda}) \label{eq:kappa:3}\text{.}
\end{equation}

Our aim it to estimate $\bfp$ by minimizing $\kappa(\bfp)$ in an alternating and parallel manner. We first estimate $\left\lbrace\bfg_{\lambda}\right\rbrace_{\lambda \in \Lambda}$, with the remaining parameters in $\bfp$ fixed as described in \fref{subsec:undirectional}, by reformulating the problem as a consensus problem. In a second step, we estimate the variables $\left\lbrace\bfm_{\lambda}, \bfd_{\lambda, 1}, \ldots, \bfd_{\lambda, Q}, \bosigma_{\lambda}^{\rmn}\right\rbrace_{\lambda \in \Lambda}$ for fixed  $\left\lbrace\bfg_{\lambda}\right\rbrace_{\lambda \in \Lambda}$, by using a sparse representation approach as described in \fref{subsec:calibrator}. During these procedures, the amount of information that needs to be exchanged between the fusion center and the compute agents is much less than the volume of data being calibrated, making this scheme computationally feasible. The overall procedure, referred to as  Parallel Calibration Algorithm (PCA), is presented in \fref{alg:proposed}. The algorithm is initialized with the true/nominal calibrator parameters and an initial guess for the antenna gains, or by default by the unit sensor gain. In the following subsections, we detail the two major alternating optimization steps of the proposed PCA.

\subsection{Undirectional Antenna Gain Estimation (\fref{alg:gain_estimation})}
\label{subsec:undirectional}
In this subsection, we describe \fref{alg:gain_estimation} of the PCA. As shown in \fref{alg:gain_estimation}, this optimization step is performed w.r.t. the undirectional gain parameters $\left\lbrace\bfg_{\lambda}\right\rbrace_{\lambda \in \Lambda}$, while the remaining parameters $\left\lbrace\bfm_{\lambda}, \bfd_{\lambda, 1}, \ldots, \bfd_{\lambda, Q}, \bosigma_{\lambda}^{\rmn}\right\rbrace_{\lambda \in \Lambda}$ of $\bfp$ are fixed. During this step, each agent calibrates the data available locally and then transfers the parameter estimates to the centralized location. At the fusion center, smoothness of the parameters across wavelength is enforced. Afterwards, this update is passed back to each compute agent. Therefore, each compute agent receives indirectly information across the whole wavelength range, thus improving the calibration.

In order to impose coherence along subbands (not along different sensors), we introduce a coherence model. Let us define $\boalpha_{k} \in \bbC^{P}$, $k = 1, \ldots, K$, such that for each sensor,
\begin{equation}
 [\bfg_{\lambda}]_{p} = \sum_{k=1}^{K} b_{\lambda, k} [\boalpha_{k}]_{p}, \forall \lambda \in \Lambda, p = 1, \ldots, P \text{.} \label{eq:10}
\end{equation}
In \fref{eq:10}, the wavelength dependence is established thanks to scalar values $b_{\lambda, k} \in \bbR$ that can be defined as polynomial terms in $\lambda$, in which the polynomial order, $K - 1$, controls the smoothness. As an example, given a reference wavelength $\lambda_0 = \frac{\rmc}{f_{0}}$, we can select $b_{\lambda, k} = \left(\frac{\lambda-\lambda_0}{\lambda_0}\right)^{1-k}$ \cite{Yat2015}. Let us denote
\begin{equation}
 \bfb_{\lambda} = \left[b_{\lambda, 1}, \ldots, b_{\lambda, K} \right]^{\sfT} \in \bbR^{K} \text{,}
\end{equation}
representing all polynomial terms and rewrite \fref{eq:10} as 
\begin{equation}
\bfg_{\lambda} = \left( \bfb_{\lambda}^{\sfT} \otimes \bfI \right) \boalpha = \bfB_{\lambda} \boalpha, \forall \lambda \in \Lambda \label{eq:def_b} \text{,}
\end{equation}
where $\bfB_{\lambda}= \left( \bfb_{\lambda}^{\sfT} \otimes \bfI \right) \in \bbR^{P \times P K}$ and $\boalpha$ is the augmented vector of hidden variables defined by
\begin{equation}
  \boalpha = \left[\boalpha_{1}^{\sfT}, \ldots, \boalpha_{K}^{\sfT} \right]^{\sfT} \in \bbC^{P K} \text{.}
\end{equation}

\begin{algorithm}[ht]
\SetAlgoRefName{1.2}
 \KwIn{$\big\lbrace \hbfR_{\lambda} \big\rbrace_{\lambda \in \Lambda}, \bfp^{[i-1]}, \eta_{\bfg}$\;}
 \textbf{Init:} set $t = 0, \bfg_{\lambda}^{[t]} = \bfg_{\lambda}^{[i-1]}$, $\bfRzero_{\lambda} = \bfA_{\bfD^{[i-1]}_{\lambda}} \bfM^{[i-1]}_{\lambda} \bfA^{\sfH}_{\bfD^{[i-1]}_{\lambda}}$\;
  \While{$\sum_{\lambda \in \Lambda} \left\| \bfg_{\lambda}^{[t-1]}-\bfg_{\lambda}^{[t]}\right\|_2 \ge \sum_{\lambda \in \Lambda} \left\| \bfg_{\lambda}^{[t]} \right\|_2 \eta_{\bfg}$}{
   \nl $t = t+1$ \;
   \nl Estimate locally $\bfg_{\lambda}^{[t]}$ with \fref{alg:step_g} \label{step_g}\;
   \nl Estimate $\boalpha^{[t]}$ by the fusion center with \fref{eq:zeta_hat}\;
   \nl Update locally $\bfy_{\lambda}^{[t]}$ with \fref{eq:y_hat}\;
  }
 \KwOut{$\big\lbrace\hbfg_{\lambda} = \bfg_{\lambda}^{[t]} \big\rbrace_{\lambda \in \Lambda}$\;}
 \caption{global estimation of $\left\lbrace\bfg_{\lambda}\right\rbrace_{\lambda \in \Lambda}$\label{alg:gain_estimation}}
\end{algorithm}

\begin{algorithm}[ht]
\SetAlgoRefName{1.2.2}
 \KwIn{$\big\lbrace \hbfR_{\lambda}, \bfRzero_{\lambda}, \bfg_{\lambda}^{[t-1]}, \bfy_{\lambda}^{[t-1]} \big\rbrace_{\lambda \in \Lambda_{z}}, \boalpha^{[t-1]}, \eta_{\bfg}$;}
 \textbf{Init:} set $t_{\lambda} = 0, \bfg_{\lambda}^{[t_{\lambda}]} = \bfg_{\lambda}^{[t-1]}$\;
  \ForEach{$\lambda \in \Lambda_{z}$}{
   \While{$\left\| \bfg_{\lambda}^{[t_{\lambda}-1]}-\bfg_{\lambda}^{[t_{\lambda}]}\right\|_2 \ge \left\| \bfg_{\lambda}^{[t_{\lambda}]} \right\|_2 \eta_{\bfg}$}{
    \nl $t_{\lambda} = t_{\lambda}+1$ \;
    \For{$p = 1, \ldots, P$}{
        \nl$\hbfr_{\lambda}^{p} = \calS_p \left(\hbfR_{\lambda} \right)$ \;
       \nl$\bfz = \calS_p \left(\bfRzero_{\lambda} \bbfG_{\lambda} \right)$ \;
       \nl$\bfz_{\boomega} = \bfz \odot \calS_p \left(\boOmega_{\lambda}\right)$ \;
       \nl$s_{p} = \sign\left([\bfg_{\lambda}^{[t_{\lambda}-1]}]_{p} - \left[\bfB_{\lambda} \boalpha\right]_{p}\right)$ \;
       \nl$[\bfg_{\lambda}^{[t_{\lambda}]}]_{p} = \frac{2\bfz_{\boomega}^{\sfH} \left(\hbfr_{\lambda}^{p} \odot \boomega \right) - s_{p} [\bbfy_{\lambda}]_{p} + \rho  \left[\bfB_{\lambda} \boalpha\right]_{p} }{2 \left(\bfz_{\boomega}^{\sfH} \bfz_{\boomega} \right) + \rho}$ \;
     }
   }
  } 
 \KwOut{$\big\lbrace\hbfg_{\lambda} = \bfg_{\lambda}^{[t_{\lambda}]}\big\rbrace_{\lambda \in \Lambda_{z}}$\;}
 \caption{local estimation of $\left\lbrace\bfg_{\lambda}\right\rbrace_{\lambda \in \Lambda_{z}}$\label{alg:step_g}}
\end{algorithm}

At this point, we distinguish between direct and parallel based estimation of $\boalpha$. Specifically:
\begin{itemize}
 \item Joint calibration leads to a direct estimation scheme of $\boalpha$ from the data. However, this requires access to the whole data by minimizing $\kappa\left(\left\lbrace\bfg_{\lambda}\right\rbrace_{\lambda \in \Lambda} \right)$ w.r.t. $\boalpha$. As explained before, this is computationally unfeasible due to the required large data volumes.
 \item To overcome this issue, we propose a parallel calibration scheme. Let us recall that $Z$ computational agents are disposed on a network (see, \fref{subsec:joint}), where the $z$-th agent, $\sfA_z$, accesses data for wavelengths $\lambda \in \Lambda_{z} \subset \Lambda$. However, we enforce consensus among all agents, by imposing the constraint $\bfg_{\lambda} = \bfB_{\lambda} \boalpha$, that each agent has to satisfy.
\end{itemize}

With this network setup, we formulate parallel calibration as
\begin{equation}
 \begin{aligned}
  \hboalpha, \left\lbrace\hbfg_{\lambda}\right\rbrace_{\lambda \in \Lambda}   = \argmin_{\boalpha, \left\lbrace\bfg_{\lambda}\right\rbrace_{\lambda \in \Lambda}} \sum_{\lambda \in \Lambda} \kappa_{\lambda}\left(\bfg_{\lambda}\right) \\
  \text{subject to~} \bfg_{\lambda} = \bfB_{\lambda} \boalpha, \forall \lambda \in \Lambda \text{,} \label{eq:pb_g}
 \end{aligned}
\end{equation}
where the cost function consists of a sum of independent cost functions, one for each subband, that are coupled through the coherence constraints which however are independent across sensors. A commonly way to solve \fref{eq:pb_g} is to consider the problem as a consensus optimization problem \cite{Boy2011} and consequently the use of the augmented Lagrangian, given by 
 \begin{align}
&L \left(\left\lbrace\bfg_{\lambda}\right\rbrace_{\lambda \in \Lambda}, \boalpha, \left\lbrace\bfy_{\lambda}\right\rbrace_{\lambda \in \Lambda} \right) =& \nonumber \\ 
&\sum_{\lambda \in \Lambda}\kappa_{\lambda}\left(\bfg_{\lambda}\right)  + \bfy_{\lambda}^{\sfH} \left(\bfg_{\lambda} - \bfB_{\lambda} \boalpha \right)  + \frac{\rho}{2} \left\| \bfg_{\lambda} - \bfB_{\lambda} \boalpha \right\|_{2}^{2} \\ 
  & = \sum_{\lambda \in \Lambda} L_{\lambda}\left(\bfg_{\lambda}, \boalpha, \bfy_{\lambda}\right) \text{,}
 \end{align}
where $\left\lbrace\bfy_{\lambda}\right\rbrace_{\lambda \in \Lambda}$ are the $J$ Lagrange multipliers and $\rho$ is the regularization term. In order to solve \fref{eq:pb_g}, we resort to the consensus ADMM \cite{Boy2011}. Let $t$ denote the local iteration counter, the values for the $[t+1]$-th iteration are updated as
\begin{align}
 \bfg_{\lambda}^{[t+1]} &= \argmin_{\bfg_{\lambda}} L_{\lambda}\left(\bfg_{\lambda}, \boalpha^{[t]}, \bfy_{\lambda}^{[t]}\right) , \lambda \in \Lambda \text{,}\label{eq:gstep} \\
 \boalpha^{[t+1]} &= \argmin_{\boalpha} \sum_{\lambda \in \Lambda} L_{\lambda}\left(\bfg_{\lambda}^{[t+1]}, \boalpha, \bfy_{\lambda}^{[t]}\right) \label{eq:zetastep} \text{,}\\
 \bfy_{\lambda}^{[t+1]} &= \bfy_{\lambda}^{[t]} + \rho \left( \bfg_{\lambda}^{[t+1]} - \bfB_{\lambda} \boalpha^{[t+1]} \right) , \lambda \in \Lambda \label{eq:y_hat} \text{,}
\end{align}
as summarized in \fref{alg:gain_estimation}. The minimization of \fref{eq:gstep} is the most computational step and is performed locally by each agent, as well as \fref{eq:y_hat}, whereas \fref{eq:zetastep} is solved by the fusion center. Procedures for obtaining  \fref{eq:gstep} and \fref{eq:zetastep} are detailed in the following.

\subsubsection{Minimization of \fref{eq:gstep}}
\label{subsubsec:gstep}

toward this aim, we follow an iterative approach based on \cite{Sal2014}, that we adapt to distributed optimization for the cost function \fref{eq:kappa:3}. We notice, firstly, that the problem is separable w.r.t. $\lambda$. Consequently, solving \fref{eq:gstep} for the mono-wavelength case is sufficient. Let us assume that $\bfg_{\lambda}$ and $\bbfg_{\lambda}$ are two independent variables. We then regard $\bbfg_{\lambda}$ as fixed and minimize $L_{\lambda}\left(\bfg_{\lambda}, \bbfg_{\lambda}, \boalpha, \bfy_{\lambda}\right) =L_{\lambda}\left(\bfg_{\lambda}\right)$ w.r.t.  $\bfg_{\lambda}$ only, and without considering the diagonal elements in the cost function \fref{eq:kappa:2} that contain the unknown noise variances $\bosigma^{\rmn}_{\lambda}$. In this case, the local cost function becomes separable w.r.t. the elements of $\bfg_{\lambda}$, hence,
\begin{equation}
 \kappa_{\lambda}(\bfg_{\lambda}) = \sum_{p=1}^{P} \kappa^{p}_{\lambda}([\bfg_{\lambda}]_{p}) \text{,} \label{eq:kappa:n}
\end{equation}
where $\kappa^{p}_{\lambda}([\bfg_{\lambda}]_{p})$ corresponds to the cost function for the $p$-th row of $\bfR_{\lambda}$, which depends only on $[\bfg_{\lambda}]_{p}$ since the remaining parameters are considered as fixed in this step. Let us define the operator $\calS_p(.)$, that converts to a vector the $p$-th row of a matrix and removes the $p$-th element of this selected vector. Further, define the vector $\hbfr_{\lambda}^{p} = \calS_p\left(\hbfR_{\lambda}\right)$ and the weighting vector $\boomega = \calS_p\left(\boOmega_{\lambda} \right)$.  We can thus write $\kappa_{\lambda}^{p}([\bfg_{\lambda}]_{p})$ in \fref{eq:kappa:n} as
\begin{equation}
 \kappa^{p}_{\lambda}([\bfg_{\lambda}]_{p}) = \left\| \left(\hbfr_{\lambda}^{p} - \bfz [\bfg_{\lambda}]_{p} \right) \odot \boomega \right\|_2^2\text{,}\label{eq:kappa:4}
\end{equation}
in which $\bfz = \calS_p \left(\bfRzero_{\lambda} \bbfG_{\lambda} \right)$ and where 
\begin{equation}
 \bfRzero_{\lambda} = \bfA_{\bfD_{\lambda}} \bfM_{\lambda} \bfA^{\sfH}_{\bfD_{\lambda}}
\end{equation}
represents the calibrator sky model. Then, we decompose the augmented Lagrangian in \fref{eq:gstep}  w.r.t. the elements of $\bfg_{\lambda}$ as
\begin{align}
 L_{\lambda}\left(\bfg_{\lambda}\right) &= \sum_{p = 1}^{P} L_{\lambda}^{p}\left([\bfg_{\lambda}]_{p}\right) \text{,}\\
 L_{\lambda}^{p}\left([\bfg_{\lambda}]_{p}\right) &= \kappa^{p}_{\lambda} \left([\bfg_{\lambda}]_{p} \right) + [\bbfy_{\lambda}]_{p} \left([\bfg_{\lambda}]_{p} - \left[\bfB_{\lambda} \boalpha\right]_{p} \right)  \nonumber\\
 & \quad + \frac{\rho}{2} \left\| [\bfg_{\lambda}]_{p} - \left[\bfB_{\lambda} \boalpha\right]_{p} \right\|_{2}^{2} \label{eq:L} \text{.}
\end{align}
By using standard inversion techniques, we set the gradient of \fref{eq:L} to zero by choosing
\begin{equation}
 [\hbfg_{\lambda}]_{p} = \frac{2\bfz_{\boomega}^{\sfH} \left(\hbfr_{\lambda}^{p} \odot \boomega \right) - s_{p} [\bbfy_{\lambda}]_{p} + \rho  \left[\bfB_{\lambda} \boalpha\right]_{p} }{2 \left(\bfz_{\boomega}^{\sfH} \bfz_{\boomega} \right) + \rho} \label{eq:g^hat} \text{,}
\end{equation}
where $\bfz_{\boomega} = \bfz \odot \calS_p \left(\boOmega_{\lambda}\right)$ and $s_{p} = \sign\left([\bfg_{\lambda}]_{p} - \left[\bfB_{\lambda} \boalpha\right]_{p}\right)$. Then, we directly update $[\bbfg_{\lambda}]_{p} = [\bar{\hbfg}_{\lambda}]_{p}$ and process in the same manner with the remaining parameters in $\bfg_{\lambda}$. This procedure is summarized  in \fref{alg:step_g} and is repeated until convergence.

\subsubsection{Minimization of \fref{eq:zetastep}}
\label{subsec:minzeta}
after gathering the estimates $\left\lbrace\bfg_{\lambda}\right\rbrace_{\lambda \in \Lambda}$, the fusion center can obtain a closed-form expression of $\boalpha$, and then, its estimated value, $\hboalpha$, is sent to all agents in the network. Specifically,
\begin{equation}
 \begin{aligned}
 \hboalpha &= \argmin_{\boalpha} \sum_{\lambda \in \Lambda} L_{\lambda}\left(\bfg_{\lambda}, \boalpha, \bfy_{\lambda}\right) \\
 &= \argmin_{\boalpha} \sum_{\lambda \in \Lambda} \bfy_{\lambda}^{\sfH} \left(\bfg_{\lambda} - \bfB_{\lambda} \boalpha \right)  \label{eq:zeta_1} + \frac{\rho}{2} \left\| \bfg_{\lambda} - \bfB_{\lambda} \boalpha \right\|_{2}^{2}\text{,}  
 \end{aligned}
\end{equation}
which leads, after some calculus, to
\begin{equation}
 \hboalpha = \left( \sum_{\lambda \in \Lambda} \rho \bfB_{\lambda}^{\sfT} \bfB_{\lambda} \right)^{\dagger} \left( \sum_{\lambda \in \Lambda} \bfB_{\lambda}^{\sfT} \left( \bfy_{\lambda} + \rho\bfg_{\lambda} \right) \right) \text{.}
\end{equation}
This above expression can be simplified by means of \fref{eq:def_b}, as
\begin{equation}
\begin{aligned}
 \hboalpha = \frac{1}{\rho} \left( \left(\sum_{\lambda \in \Lambda} \bfb_{\lambda} \bfb_{\lambda}^{\sfT} \right)^{\dagger} \otimes \bfI \right)
 \left( \sum_{\lambda\in \Lambda} \bfb_{\lambda} \circ \left( \bfy_{\lambda} + \rho\bfg_{\lambda} \right) \right) \label{eq:zetaint} \text{.}
\end{aligned} 
\end{equation}
For obtaining $\hboalpha$ from \fref{eq:zetaint}, we request $J \geq K$, i.e., accessing to data for at least $K$ wavelengths, which is supposed satisfied since, e.g., for the LOFAR, the signal is typically divided into 512 subbands while usually a low order polynomial is used. Finally, denoting
\begin{align}
 \bfT =& \frac{1}{\rho} \left(\sum_{\lambda\in \Lambda} \bfb_{\lambda} \bfb_{\lambda}^{\sfT} \right)^{\dagger} \text{,}\\
 \bfu =& \vect \left(\bfU \right) = \sum_{\lambda \in \Lambda} \bfb_{\lambda} \circ \left( \bfy_{\lambda} + \rho\bfg_{\lambda} \right) 
\end{align}
and by use of the Kronecker product property $\vect(\bfA \bfB \bfC) = \left( \bfC^{\sfT} \otimes \bfA \right) \vect(\bfB)$, \fref{eq:zetastep} is reduced to the following compact analytical expression,
\begin{equation}
\hboalpha = \vect \left(\bfI \bfU \bfT^{\sfT} \right) = \vect \left(\bfU \bfT^{\sfT} \right) \label{eq:zeta_hat}\text{.}
\end{equation}

\subsection{Directional Parameter and Noise Power Estimation (\fref{alg:iht})}
\label{subsec:calibrator}
In this subsection, we describe \fref{alg:iht} of the PCA dedicated to the estimation of the directional parameters and noise powers $\left\lbrace\bfm_{\lambda}, \bfd_{\lambda, 1}, \ldots, \bfd_{\lambda, Q}, \bosigma_{\lambda}^{\rmn}\right\rbrace_{\lambda \in \Lambda}$ for fixed  $\left\lbrace\bfg_{\lambda}\right\rbrace_{\lambda \in \Lambda}$, which is based mainly on a sparse representation framework.

Assuming that the calibration sources are well separated, which is common in radio astronomy \cite{Wijthe, Ven2013}, we consider in the remainder of this paper that for every wavelength: i) each apparent calibration source lies in a sector of displacements around its nominal location; and ii) the displacement sectors of different calibration sources are not overlapping. Consequently, each dictionary shall represent the displacement set corresponding to its source. Towards this aim, let us define $J Q$ dictionaries of steering vectors, $\tbfA_{q, \lambda}$, for $q=1,\ldots, Q, \lambda \in \Lambda$, as
\begin{equation}
\begin{aligned}
 \tbfA_{q, \lambda} = \Big[ \bfa_{\lambda}(d_{l, \lambda, q}^{1}, d_{m, \lambda, q}^{1}), \ldots ,\bfa_{\lambda}(d_{l, \lambda, q}^{1}, d_{m, \lambda, q}^{N_q^{m}}), \\
 \bfa_{\lambda}(d_{l, \lambda, q}^{2}, d_{m, \lambda, q}^{1}),\ldots, \bfa_{\lambda}(d_{l, \lambda, q}^{N_q^{l}}, d_{m, \lambda, q}^{N_q^{m}}) \Big] \in \bbC^{P \times N_q^{l}N_q^{m}} \text{,}
\end{aligned}
 \end{equation}
which contain  $N_q^{l}N_q^{m}$ steering vectors, centered around the true/nominal direction of the $q$-th calibrator, namely $\bfd_{q}^{\TRUE}$, with resolution $\left(\Delta l_{\lambda}^{q},\Delta m_{\lambda}^{q} \right)$ and $N_q^{l} \gg 1, N_q^{m} \gg 1$. Let us recall that the direction shifts are proportional to $\lambda^{2}$ \cite{Coh2009} (see, \fref{subsec:wavelength}). Consequently, we impose the same behavior w.r.t. the wavelength in the step resolutions, i.e., $\Delta l_{\lambda}^{q}  \propto \lambda^{2}, \Delta m_{\lambda}^{q}  \propto \lambda^{2}$, by scaling them around $\lambda_0$ as
\begin{align}
 \Delta l_{\lambda}^{q} &= \left(\frac{\lambda}{\lambda_0}\right)^{2} \Delta l_{\lambda_0}^{q} \text{,}\\
 \Delta m_{\lambda}^{q} &= \left(\frac{\lambda}{\lambda_0}\right)^{2} \Delta m_{\lambda_0}^{q}\text{.}
\end{align}
 These dictionary steering matrices are gathered in 
\begin{equation}
 \tbfA_{\lambda} = \left[ \tbfA_{1, \lambda}, \ldots ,\tbfA_{Q, \lambda} \right] \in \bbC^{P \times N_g}, \lambda \in \Lambda \text{,}
\end{equation}
with $N_g = \sum_{q=1}^{Q} N_q^{l}N_q^{m} $ denoting the total number of directions on the grid. 

\begin{algorithm}[ht]
\SetAlgoRefName{1.3}
 \KwIn{$\big\lbrace\hbfR_{\lambda}, \bfg_{\lambda}^{[i]}\big\rbrace_{\lambda \in \Lambda}, \bfp^{[i-1]}, \eta_{\tbfm}$\;}
 \textbf{Init:} set $k = 0$, $\bfg_{\lambda}^{[k]} = \bfg_{\lambda}^{[i]}$, $\bfM_{\lambda}^{[k]} = \bfM^{[i-1]}_{\lambda}$, $\bfD_{\lambda}^{[k]} = \bfD^{[i-1]}_{\lambda}, \bosigma_{\lambda}^{\rmn [k]} = \bosigma_{\lambda}^{\rmn[i-1]}$\;
 \While{$\left\| \tbfm^{[k-1]}- \tbfm^{[k]} \right\|_2 \ge \left\| \tbfm^{[k]} \right\|_2 \eta_{\tbfm}$}{
 \nl $k = k+1$\;
 \For{$q=1,\ldots, Q$}{ 
  \ForEach{$\sfA_z, z = 1, \ldots, Z$}{
   \ForEach{$\lambda \in \Lambda_{z}$}{
    \nl Calculate locally $\chbfr_{\lambda}^{q} = \chbfr_{\lambda} - \sum_{q'=1}^{q-1} \cbfV_{\lambda}^{q'} \tbfm_{q'}^{[k]} - \sum_{q'=q+1}^{Q} \cbfV_{\lambda}^{q'} \tbfm_{q'}^{[k-1]}$\;
    }
   \nl Send $\sum_{\lambda \in \Lambda_{z}} \cbfV_{\lambda}^{q \sfT} \chbfr_{\lambda}^{q}$ to the fusion center\;
   }
  \nl Hard thresholding by the fusion center $\tbfm_q^{[k]} = \tau_{q} \calH_1 \left(  \sum_{\lambda \in \Lambda} \cbfV_{\lambda}^{q \sfT} \chbfr_{\lambda}^{q} \right)$\;
  \nl The fusion center communicates the non-zero element of $\tbfm_{\ q}^{[k]}$ and $\bfd_{\lambda_{0}, q}^{[k]}$\;
  }
 }
 \nl Estimate locally $\bosigma_{\lambda}^{\rmn}$ with \fref{eq:sigman:2}\;
 \caption{estimation of $\tbfm$ and $\left\lbrace \bosigma_{\lambda}^{\rmn}\right\rbrace_{\lambda \in \Lambda}$\label{alg:iht}}
 \KwOut{$\big\lbrace\hbfm_{\lambda}, \hbfd_{\lambda, 1}, \ldots, \hbfd_{\lambda, Q}, \hbosigma_{\lambda}^{\rmn}\big\rbrace_{\lambda \in \Lambda}$ \;}
\end{algorithm}

We define then $J$ vectors, $\{\tbfm_{\lambda}\}_{\lambda \in \Lambda}$, as
\begin{equation}
 \tbfm_{\lambda} = \left[ \tbfm_{\lambda, 1}^{\sfT}, \ldots ,\tbfm_{\lambda, Q}^{\sfT} \right]^{\sfT} \in \bbR^{N_g}, \lambda \in \Lambda \text{,}
\end{equation}
which contains the squared directional gains of all calibrators, where $\tbfm_{\lambda, q}$ is the sparse vector associated with $\tbfA_{q, \lambda}$. Due to the previous assumption of non-overlapping displacement sectors, each $\tbfm_{\lambda, q}$ is exactly $1$-sparse, i.e., $\left\| \tbfm_{\lambda, q} \right\|_0=1$, for $q = 1, \ldots, Q,  \forall \lambda \in \Lambda$. Since the shift resolution in the dictionaries is made proportional to $\lambda^{2}$, the support of $\tbfm_{\lambda, q}$ is independent of $\lambda$. To go further, we exploit that $\boGamma_{\lambda} \propto \lambda^{-1}$ (see, \fref{subsec:wavelength}) in order to estimate a unique sparse vector for the all wavelengths, namely $\tbfm$. More precisely, under this assumption, we define $\tbfm = \left[\tbfm_{1}^{\sfT}, \ldots, \tbfm_{Q}^{\sfT} \right]^{\sfT}$ as
\begin{equation}
 \tbfm_{\lambda} = \left( \frac{\lambda_{0}}{\lambda} \right)^{2} \tbfm ,\forall \lambda \in \Lambda \text{,}
\end{equation}
which can be modified for another given behavior of $\boGamma_{\lambda}$.

Using \fref{eq:cov}, the covariance model can be rewritten as
\begin{equation}
\begin{aligned}
  \bfR_{\lambda} &= \tbfE_{\lambda} \tbfM \tbfE_{\lambda}^{\sfH} +\bfR^{\OUT}_{\lambda} + \boSigman_{\lambda} \text{,}
\end{aligned}
\end{equation}
in which $\tbfM = \diag(\tbfm)$ and
\begin{equation}
 \tbfE_{\lambda} = \frac{\lambda_{0}}{\lambda} \bfG_{\lambda} \tbfA_{\lambda} \boSigma^{\frac{1}{2}}_{\lambda} \text{.}
\end{equation}

Let us then define
\begin{align}
 \bfV_{\lambda} &= \left ( \boSigma^{\rmn}_{\lambda} \right )^{-\frac{1}{2}} \btbfE_{\lambda} \circ \left ( \boSigma^{\rmn}_{\lambda} \right )^{-\frac{1}{2}} \tbfE_{\lambda} \text{,} \\
 \bfN_{\lambda} &= \boSigma^{\rmn-\frac{1}{2}}_{\lambda} \circ \boSigma^{\rmn-\frac{1}{2}}_{\lambda}  \text{,}\\
 \hbfr_{\lambda} &= \vect\left(\hbfR_{\lambda} \odot \boSigma^{\rmn}_{\lambda}\right)\text{.}
\end{align}
Thus, we formulate the minimization problem as 
\begin{equation}
\begin{aligned}
&\htbfm, \left\lbrace \hbosigma_{\lambda}^{\rmn} \right\rbrace_{\lambda \in \Lambda}  = \argmin_{\tbfm, \left\lbrace\bosigma^{\rmn}_{\lambda}\right\rbrace_{\lambda \in \Lambda}} \sum_{\lambda \in \Lambda} \left\| \hbfr_{\lambda} - \bfV_{\lambda} \tbfm - \bfN_{\lambda} \bosigma^{\rmn}_{\lambda} \right\|^2_2 \\
  &\text{subject to~} \tbfm \succeq \bfzero, \bosigma^{\rmn}_{\lambda} \succeq \bfzero, \forall \lambda \in \Lambda \\
  &\hspace*{1.6cm}\left\| \tbfm_{q} \right\|_0 = 1 \text{~for~} q=1,\ldots, Q \text{.}
\end{aligned} \label{eq:mu_1}
\end{equation}
Since the $p$-th element of $\bosigma^{\rmn}_{\lambda}$, $[\bosigma_{\lambda}^{\rmn}]_{p}$, is only present in the $p$-th diagonal term of $\bfR_{\lambda}$, ignoring this term does not affect the estimation of $\tbfm$ if $[\bosigma^{\rmn}_{\lambda}]_{p}> 0$ holds. Consequently, we estimate $\bosigma^{\rmn}_{\lambda}$ after $\tbfm$.

For this purpose, let us denote $\chbfr_{\lambda}$ and $\cbfV_{\lambda}$, that refer, respectively, to $\hbfr_{\lambda}$ and $\bfV_{\lambda}$ without their elements corresponding to the diagonal of $\bfR_{\lambda}$. We define also
\begin{equation}
 h_{\lambda}(\tbfm) = \left\| \chbfr_{\lambda} - \cbfV_{\lambda} \tbfm \right\|^2_2, \lambda \in \Lambda \text{,}
\end{equation}
to obtain the solution of $\tbfm$ in \fref{eq:mu_1} after supposing $\bosigma^{\rmn}_{\lambda} \succ \bfzero$, as
\begin{equation}
\begin{aligned}
&\htbfm  = \argmin_{\tbfm}  \sum_{\lambda \in \Lambda} h_{\lambda}(\tbfm)\\
  &\text{subject to~} \tbfm \succeq \bfzero,\\
  &\hspace*{1.6cm}\left\| \tbfm_{q} \right\|_0 = 1 \text{~for~} q=1,\ldots, Q \text{,} \label{eq:min1}
\end{aligned}
\end{equation}
which is used in \fref{alg:iht}. 

To consider the $l_0$ constraints in \fref{eq:min1}, which are non-convex and NP-hard to solve \cite{Can2006}, we choose the Distributed Iterative Hard Thresholding \cite{Pat2013, Pat2014, Han2015}, which is based on the Iterative Hard Thresholding \cite{Blu2008}. This greedy algorithm consists of a projected gradient descend direction algorithm and offers strong theoretical guarantees that have been succefully employed in the DoA estimation context \cite{Oll2014, Oll2015_1}. Particularly, when the grid is fine and the columns of $\tbfA_{q, \lambda}$ are strongly coherent, we can guarantee that each $\tbfm_q$ obtained from \fref{eq:min1} is exactly $1$-sparse. Thus, using the Coordinate Descent algorithm \cite{Fri2007} to minimize \fref{eq:min1}, we obtain an analytic solution for each sub-problem and the sparsity of the desired minimizer $\tbfm$ reduces the computational complexity. Each step involves the hard thresholding operator $\calH_s(.)$, that keeps the $s$-largest components of a vector and sets the remaining entries equal to zero, thus, it automatically satisfies both constraints of sparsity and positivity. We can allow a step size factor $\tau^{k}_{q}$ that depends on $\tbfm_{q}$ and the $k$-th iteration, by use of the Normalized Iterative Hard Thresholding \cite{Blu2010}. Let us take benefit from the $1$-sparsity of $\tbfm_q$ in order to choose $\tau^{k}_{q}$: firstly, we define its associated residual as
\begin{equation}
  \chbfr_{\lambda}^{q} = \chbfr_{\lambda} - \sum_{\substack{q' = 1 \\ q' \ne q}}^{Q} \cbfV_{\lambda}^{q'} \tbfm_{q'} \text{.}
\end{equation}
Secondly, we obtain the update for the first iteration as
\begin{equation}
 \tbfm_{q}^{[1]} = \calH_{1} \left( \tbfm_{q}^{[0]} +  \tau^{1}_{q} \sum_{\lambda \in \Lambda}  \cbfV_{\lambda}^{q \sfT} \left(\chbfr_{\lambda}^{q} - \cbfV_{\lambda}^{q} \tbfm_{q}^{[0]}\right) \right) \text{.}
\end{equation}
Then, we consider the initialization $\tbfm_{q}^{[0]} = \bfzero$ and note that the dictionary of $\tbfm_{q}$ is given by 
\begin{equation}
 \cbfV^{q} = \left[ \cbfV_{\lambda_{1}}^{q \sfT}, \ldots, \cbfV_{\lambda_{J}}^{q \sfT} \right]^{\sfT} \text{,}
\end{equation}
whose columns have the same norm. Consequently, by choosing
\begin{equation}
\tau^{1}_{q} = \tau_{q} = \frac{1}{\left\| \cbfv^{q} \right\|^{2}_{2}} = \frac{1}{\sum_{\lambda \in \Lambda} \left\| \cbfv_{\lambda}^{q} \right\|^{2}_{2}} \text{,}
\end{equation}
where $\cbfv^{q}$ and $\cbfv^{q}_{\lambda}$ are, respectively, any column of $\cbfV^{q}$ and $\cbfV^{q}_{\lambda}$, we obtain directly the solution for $\tbfm_{q}$ after the first iteration.

In the network, the $z$-th agent, $\sfA_{z}$, accesses only $\{h_{\lambda}(.)\}_{\lambda \in \Lambda_{z}}$. In order to estimate $\tbfm_q$ and then deduce its DoA and $\bfm_{\lambda, q}$, each agent $\sfA_z$ can calculate the values $\sum_{\lambda \in \Lambda_{z}} \cbfV_{\lambda}^{q \sfT} \chbfr_{\lambda}^{q}$ and send them to the fusion center, which processes for thresholding. Then, the fusion center transmits only the non-zero value of $\tbfm_{q}$ and its corresponding direction $\bfd_{\lambda_{0}, q}$, as drawn in \Fref{fig:dessin}. Benefiting from the positivity of $\tbfm_{q}$, we are able to implement the procedure in \cite{Pat2013}, that solves a top-$K$ problem. Thus, the $Z$ agents can send only a fraction of the estimates to the fusion center for saving transmission cost. This procedure is not described in \fref{alg:iht} for convenience, since it only improves the communication efficiency.

Afterward, the estimation of $\left\lbrace\bosigma^{\rmn}_{\lambda}\right\rbrace_{\lambda \in \Lambda}$ is performed locally, without the need of transmitting the estimated values. Firstly, note that without considering outliers, i.e., $\bfR^{\OUT}_{\lambda} \approx \bfzero$, the estimation of $\bosigma^{\rmn}_{\lambda}$ is given by
\begin{equation}
 \cbosigma_{\lambda}^{\rmn} = \vectdiag\left(\hbfR_{\lambda}- \hbfG_{\lambda} \hbfRzero_{\lambda} \hbfG^{\sfH}_{\lambda} \right) \text{,}
\end{equation}
since we assume independence of $\bosigma_{\lambda}^{\rmn}$ across wavelength.
Secondly, we remove the bias introduced by the outliers as follows: we calculate the power
\begin{equation}
 \sigma_{\lambda}^{r} = \frac{\bfa^{\sfH}_{\lambda}(\bfd_{\lambda, r}) \left( \hbfR_{\lambda} - \hbfG_{\lambda} \hbfRzero_{\lambda} \hbfG^{\sfH}_{\lambda} \right) \bfa_{\lambda}(\bfd_{\lambda, r})}{\left\|\bfa_{\lambda}(\bfd_{\lambda, r})\right\|_2^2 } \label{eq:sigman:1}
\end{equation}
of the residual sample covariance matrix for a random direction $\bfd_{\lambda, r}$, where no source is supposed to be present. We then approximate $\bfa^{\sfH}_{\lambda}(\bfd_{\lambda, r}) \bfa_{\lambda}(\bfd_{\lambda, q}) \approx 0$ for any $\bfd_{\lambda, r} \ne \bfd_{\lambda, q}$, which yields $\sigma_{\lambda}^{r}$ as the sum of the sensor noise powers \cite{Les2000, Ven2013}. By imposing $\sum_{p=1}^{P} \left[\bosigma_{\lambda}^{\rmn}\right]_{p} = \sigma_{\lambda}^{r}$, the new unbiased solution is given by
\begin{equation}
 \hbosigma_{\lambda}^{\rmn} = \cbosigma_{\lambda}^{\rmn} +\frac{1}{P} \left( \sigma_{\lambda}^{r} - \bfun_{P \times 1}^{\sfT} \cbosigma_{\lambda}^{\rmn} \right) \bfun_{P \times 1} \text{,}
\label{eq:sigman:2}
\end{equation}
that concludes \fref{alg:iht}.

\section{Cramér-Rao Bound}
\label{sec:cr_bound}
The Cramér-Rao Bound (CRB) expresses a lower bound on the variance of the estimation error of a deterministic vector parameter for an unbiased estimator \cite{Kay1993, Sto2005}. In this section, after obtaining the CRB for the mono-wavelength scenario, we define the unconstrained CRB in the multi-wavelength scenario and finally take into account the dependence across wavelength (see, \fref{subsec:wavelength}) to obtain the constrained CRB that corresponds to the data model \fref{eq:cov}.

Let us consider the mono-wavelength scenario and stacking the unknown parameters in
\begin{equation}
\botheta_{\lambda} = \big[\Re(\bfg_{\lambda}^{\sfT}), \Im(\bfg_{\lambda}^{\sfT}), \bfm_{\lambda}^{\sfT}, \bfm_{\lambda}^{\OUT \sfT}, \bfd_{l, \lambda}^{\sfT}, \bfd_{l, \lambda}^{\OUT \sfT},  \bfd_{m, \lambda}^{\sfT}, \bfd_{m, \lambda}^{\OUT \sfT}, \bosigma^{\rmn \sfT}_{\lambda} \big]^{\sfT} \text{,}
\end{equation}
where $\boGamma_{\lambda}^{\OUT} \boGamma_{\lambda}^{\OUT \sfH} =  \diag\left(\bfm_{\lambda}^{\OUT}\right)$, $\bfd_{l, \lambda} = \left[d_{l, \lambda, 1}, \ldots, d_{l, \lambda, Q} \right]^{\sfT}$, $\bfd_{l, \lambda}^{\OUT} = [d_{l, \lambda, 1}^{\OUT}, \ldots, d_{l, \lambda, Q^{\OUT}}^{\OUT}]^{\sfT}$, $\bfd_{m, \lambda} = [d_{m, \lambda, 1}, \ldots, d_{m, \lambda, Q} ]^{\sfT}$ and $\bfd_{m, \lambda}^{\OUT} = [d_{m, \lambda, 1}^{\OUT}, \ldots, d_{m, \lambda, Q^{\OUT}}^{\OUT} ]^{\sfT}$. We obtain its associated CRB, $\bfC_{\lambda}$, after straightforward adaptations from [6, Chapter 4]. Then, for the multi-wavelength scenario, we gather the unknown parameters in a vector $ \botheta = [\botheta_{\lambda_{1}}^{\sfT}, \ldots, \botheta_{\lambda_{J}}^{\sfT}]^{\sfT}$, suppose that the signals are i.i.d. across wavelength and ignore the constraints for the parameters. Consequently, we obtain the unconstrained CRB, $\tbfC$, as
\begin{equation}
\tbfC =
 \begin{bmatrix}
   \bfC_{\lambda_{1}} & \bfzero  & \bfzero \\
   \bfzero & \ddots & \bfzero \\
   \bfzero & \bfzero & \bfC_{\lambda_{J}}
\end{bmatrix} \text{.}
\end{equation}

From $\tbfC$, we obtain the CRB corresponding to the data model, $\bfC$, as \cite{Mar1993}
\begin{equation}
\bfC = \tbfC - \tbfC \boPsi^{\sfT} (\boPsi \tbfC \boPsi^{\sfT})^{-1} \boPsi \tbfC^{\sfT} \text{,}
\end{equation}
where $\boPsi$ is the gradient matrix of the constraints, given by
\begin{equation}
 \boPsi = \left[\boPsi_{\bfg_{\lambda}}^{\sfT}, \boPsi_{\bfm_{\lambda}}^{\sfT}, \boPsi_{\bfm_{\lambda}^{\OUT}}^{\sfT}, \boPsi_{\bfD_{\lambda}}^{\sfT}, \boPsi_{\bfD_{\lambda}^{\OUT}}^{\sfT} \right]^{\sfT} \text{,}
\end{equation}
in which the constraints on $\bfg_{\lambda}, \bfm_{\lambda}, \bfm_{\lambda}^{\OUT}, \bfD_{\lambda}$ and $\bfD_{\lambda}^{\OUT}, \lambda \in \Lambda$, are represented in  $\boPsi_{\bfg_{\lambda}}, \boPsi_{\bfm_{\lambda}}, \boPsi_{\bfm_{\lambda}^{\OUT}}, \boPsi_{\bfD_{\lambda}}$ and $\boPsi_{\bfD_{\lambda}^{\OUT}}$, respectively. Since $\bfm_{\lambda} \propto \lambda^{-2}$, we have
\begin{equation}
 \lambda_{1}^{2} \bfm_{\lambda_{1}} = \lambda_{2}^{2} \bfm_{\lambda_{2}} = \ldots = \lambda_{J}^{2} \bfm_{\lambda_{J}} \text{,}
\end{equation}
leading to
\begin{equation}
 \boPsi_{\bfm_{\lambda}} =
  \begin{bmatrix}
   \lambda_{1}^{2}\bfI &  -\lambda_{2}^{2}\bfI & \bfzero & \ldots & \bfzero \\
   \lambda_{1}^{2}\bfI & \bfzero & \ddots & & \bfzero \\
   \vdots & \vdots & & \ddots & \vdots \\
   \lambda_{1}^{2}\bfI & \bfzero & \ldots & \bfzero & -\lambda_{J}^{2}\bfI\\
\end{bmatrix} 
\end{equation}
and we add zeros for the indices corresponding to the remaining parameters in $\botheta$. $\boPsi_{\bfm_{\lambda}^{\OUT}}$ is obtained in the same way. On the other hand, in order to derive $\boPsi_{\bfD_{\lambda}}$, we make use of the following constraints 
\begin{align}
 \lambda_{1}^{-2} d_{l, q, \lambda_{1}} &= \lambda_{2}^{-2} d_{l, q, \lambda_{2}} = \ldots = \lambda_{J}^{-2} d_{l, q, \lambda_{J}} \text{,} \\
 \lambda_{1}^{-2} d_{m, q, \lambda_{1}} &= \lambda_{2}^{-2} d_{m, q, \lambda_{2}} = \ldots = \lambda_{J}^{-2} d_{m, q, \lambda_{J}} \text{,}
\end{align}
leading to
\begin{equation}
 \boPsi^{l}_{q} =
  \begin{bmatrix}
   \lambda_{1}^{-2} &  -\lambda_{2}^{-2} & 0 & \ldots & 0 \\
   \lambda_{1}^{-2} & 0 & \ddots & & 0\\
   \vdots & \vdots & & \ddots & \vdots \\
   \lambda_{1}^{-2} & 0 & \ldots & 0 & -\lambda_{J}^{-2}\\
\end{bmatrix} 
\end{equation}
and we add zeros for the indices corresponding to the remaining parameters in $\botheta$ and process in a same way for $d_{m}$ to obtain $\boPsi^{m}_{q}$. Thus, $\boPsi_{\bfD_{\lambda}}$ is given by
\begin{equation}
 \boPsi_{\bfD_{\lambda}} = \left[ \boPsi^{l \sfT}_{1}, \ldots, \boPsi^{l \sfT}_{Q}, \boPsi^{m \sfT}_{1}, \ldots, \boPsi^{m \sfT}_{Q} \right]^{\sfT} \text{,}
\end{equation}
and we derive in the same way $\boPsi_{\bfD_{\lambda}^{\OUT}}$. Finally, we consider the following constraint
\begin{equation}
 \bfg_{\lambda} = \bfB_{\lambda} \boalpha, \forall \lambda \in \Lambda \text{,}
\end{equation}
that reduces the degree of freedom of $\{\bfg_{\lambda}\}_{\lambda \in \Lambda}$ from $JP$ to $KP$, i.e., we add $(J-K)P$ constraints. Let us define
\begin{align}
 \bfg_{K} &= [\bfg_{\lambda_{1}}^{\sfT}, \ldots, \bfg_{\lambda_{K}}^{\sfT} ]^{\sfT} \text{,} \\
 \bfB_{K} &=   \left[\bfB_{\lambda_{1}}^{\sfT}, \ldots, \bfB_{\lambda_{K}}^{\sfT}\right]^{\sfT} \text{.}
\end{align}
Thus,
\begin{equation}
 \bfg_{\lambda} = \bfB_{\lambda} \bfB_{K}^{-1} \bfg_{K}, \lambda = \lambda_{K+1}, \ldots, \lambda_{J} \text{,}
\end{equation}
leading to
\begin{equation}
 \boPsi_{\bfg_{\lambda}} =
  \begin{bmatrix}
   \bfB_{\lambda_{K+1}} \bfB_{K}^{-1} &  -\bfI & \bfzero & \ldots & \bfzero \\
   \bfB_{\lambda_{K+2}} \bfB_{K}^{-1} & \bfzero & \ddots & & \bfzero \\
   \vdots & \vdots & & \ddots & \vdots \\
   \bfB_{\lambda_{J}} \bfB_{K}^{-1} & \bfzero & \ldots & \bfzero & -\bfI\\
\end{bmatrix}
\end{equation}
and we add zeros for the indexes corresponding to the remaining parameters in $\botheta$, which concludes our derivation of the constrained CRB.

\section{Simulations}
\label{sec:simulations}

The proposed method is evaluated in realistic situations, with similar sensor locations of LOFAR’s Initial Test Station \cite{Wij2004}, with typical parameter values commonly used in radio astronomy applications \cite{Wij2010, Wijthe, Ven2013}. In order to analyze the estimation of $\{\bfg_{\lambda}\}_{\lambda \in \Lambda}$, we first focus on \fref{alg:gain_estimation} and \fref{alg:step_g} and then show results for the PCA.

 \begin{figure}[ht]
  \centering
    \begin{tikzpicture}[]
    \begin{axis}[only marks, height=5.0cm, width=5.0cm, xlabel={$x$ (m)},ylabel={$y$ (m)},
       ylabel style={xshift=-0ex,yshift=-3ex},
    xlabel style={xshift=-0ex,yshift=+1ex}]
     \addplot[mark=+] table[x=x,y=y] {sensor_location.txt};
    \end{axis}
  \end{tikzpicture}
  \caption{LOFAR’s Initial Test Station antenna locations \cite{Wij2004}.\label{fig:its}}
 \end{figure}
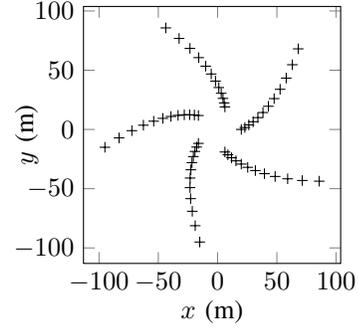
 
 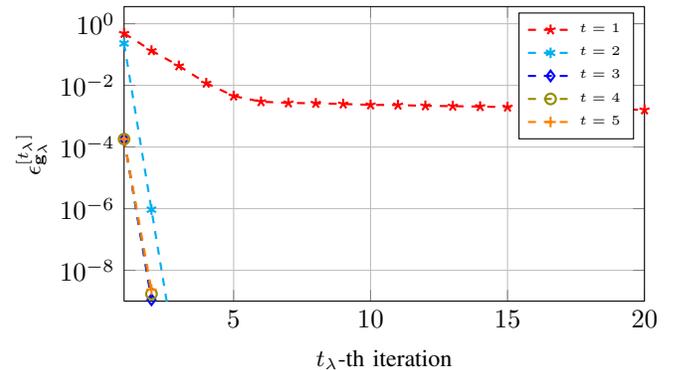
\begin{figure}[hb]
 \centering
 \begin{tikzpicture}[]
  \begin{semilogyaxis}[height=5.5cm, width=8.5cm, enlarge x limits=false, xlabel= $t_{\lambda}$-th iteration, ylabel= $\epsilon_{\bfg_{\lambda}}^{[t_{\lambda}]}$, legend style={font=\tiny}, legend columns=1, grid=both, ymin = 0.000000001,
   ytick={1, 0.01, 0.0001, 0.000001, 0.00000001},
   label style={font=\small}, ticks=both, legend entries={{$t = 1$}, {$t = 2$}, {$t = 3$}, {$t = 4$}, {$t = 5$}},
   mark options={solid}]
   \addplot+[dashed, mark=star, draw = red, thick] table[x=iter,y=t1] {t_lambda_residual.txt};
   \addplot+[dashed, mark=asterisk, draw = cyan, thick] table[x=iter,y=t2] {t_lambda_residual.txt};
   \addplot+[dashed, mark=diamond, draw = blue, thick] table[x=iter,y=t3] {t_lambda_residual.txt};
   \addplot+[dashed, mark=o, draw = olive, thick] table[x=iter,y=t4] {t_lambda_residual.txt};
   \addplot+[dashed, mark=+, draw = orange, thick] table[x=iter,y=t5] {t_lambda_residual.txt};
   \end{semilogyaxis}
  \end{tikzpicture}
 \caption{$\bfg_{\lambda}$-residual, $\epsilon_{\bfg_{\lambda}}$, as function of the iteration number $t_{\lambda}$ of \fref{alg:step_g}, for different values of the $t$-th iteration of \fref{alg:gain_estimation}\label{fig:results_g_5}.}
\end{figure}

\subsection{Data Setup}
\begin{figure}[ht]
\centering
  \begin{tikzpicture}[]
   \begin{semilogyaxis}[height=5.5cm, width=8.5cm, enlarge x limits=false, xlabel= $t$-th iteration, ylabel= {$\epsilon_r^{[t]}$}, legend style={font=\tiny}, legend columns=1,
   label style={font=\small}, grid=both, ticks=both, legend entries={{$K=2, \rho=5\cdot10^3 P$}, {$K=3, \rho=5\cdot10^2 P$}, {$K=3, \rho=5\cdot10^3 P$}, {$K=3, \rho=5\cdot10^4 P$} , {$K=4, \rho=5\cdot10^3 P$}}, 
   ytick={1, 0.1, 0.01 ,0.001, 0.0001 , 0.00001, 0.000001}, 
   ,mark options={solid}]
    \addplot+[dashed, mark=star, draw = red, thick] table[x=iter,y=r1] {primal_residual.txt};
    \addplot+[dashed, mark=asterisk, draw = cyan, thick] table[x=iter,y=r2] {primal_residual.txt};
    \addplot+[dashed, mark=diamond, draw = blue, thick] table[x=iter,y=r3] {primal_residual.txt};
    \addplot+[dashed, mark=o, draw = olive, thick] table[x=iter,y=r4] {primal_residual.txt};
    \addplot+[dashed, mark=+ , draw = orange, thick] table[x=iter,y=r5] {primal_residual.txt};
   \end{semilogyaxis}
  \end{tikzpicture}
 \caption{Primal residual, $\epsilon_r$, as function of the iteration number $t$ of \fref{alg:gain_estimation}, for smoothing polynomial terms $K = 2,3,4$ and regularization term $\rho = 5\cdot10^2 P, 5\cdot10^3 P, 5\cdot10^4 P$. \label{fig:results_g_1}}
\end{figure}
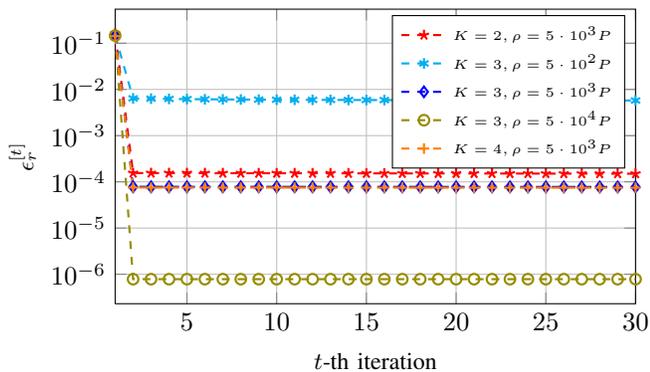

The antenna locations correspond to the LOFAR’s Initial Test Station \cite{Wij2004}, with $P = 60$ antennas disposed in a five-armed spiral, as shown in \Fref{fig:its}. We assume a sky model with $\lambda_{0} = \SI{10}{m}$ ($f_0 = \SI{30}{MHz}$) consisting of $Q=2$ strong calibration sources and $Q^{\OUT}=8$ weak non-calibration sources, provided from the ten strongest sources in the table of \cite{Ben1962}. The total power of these sources is assumed to be 1\% of the total antenna noise power, a typical scenario for radio interferometers \cite{Wijthe}. We consider data taken at $J = 13$ wavelengths, uniformly spaced in frequency from \SI{7.1}{m} to \SI{16.5}{m}. We create $\{\bfg_{\lambda}\}_{\lambda \in \Lambda}$ by using a polynomial of order $K^{\TRUE}=3$, with $b_{\lambda, k} = \left(\frac{\lambda-\lambda_0}{\lambda_0}\right)^{1-k}$, given as one realization sample from $\calC\calN\left(\bfun, \left(\sigma^{\boalpha}_{\bbR} +\rmj\sigma^{\boalpha}_{\bbC}\right) \bfI\right)$ with $\sigma^{\boalpha}_{\bbR} = \sigma^{\boalpha}_{\bbC} = 0.25$ and we consider $\bfg_{\lambda}^{[0]} = \bfun$ as an initialization. To initialize \fref{alg:gain_estimation}, we consider the regularization parameter $\rho$ as null during the first estimation of $\bfg_{\lambda}$, i.e., the first estimation of $\bfg_{\lambda}$ is done without enforcing smoothness.  We generate the shifts for $\left(l_{q, \lambda_{0}}, m_{q, \lambda_{0}}\right), q = 1, \ldots, Q$ and the diagonal of $\boGamma_{\lambda_{0}}$ with one realization sample from $\calU\left( \left(l_{q}^{\TRUE}, m_{q}^{\TRUE}\right), \sigma_{\lambda_{0}}^{\bfD} \bfun \right)$ and $\calC\calU\left(\bfun, \left(\sigma^{\boGamma}_{\bbR} +\rmj\sigma^{\boGamma}_{\bbC}\right) \bfun\right)$, respectively, with $\sigma_{\lambda_{0}}^{\bfD} = 10^{-1}/\sqrt{3}, \sigma_{\bbR}^{\boGamma} = \sigma_{\bbC}^{\boGamma} = 1/\sqrt{60}$, and initialize with $\boGamma_{\lambda} = \bfI$. Data are produced via the signal model given in \fref{eq:signal}, in order to obtain the sample covariance matrices \fref{eq:sample_cov}.

\subsection{Results}

\subsubsection{Results for the estimation of $\{\bfg_{\lambda}\}_{\lambda \in \Lambda}$}
we illustrate here the convergence and the performances of both \fref{alg:gain_estimation} and \fref{alg:step_g}. In order to analyse convergence, we define the $\bfg_{\lambda}$-residual, $\epsilon_{\bfg_{\lambda}}$, the primal residual, $\epsilon_r$, and the dual residual, $\epsilon_d$, as
\begin{align}
\epsilon_{\bfg_{\lambda}}^{[t_{\lambda}]} &= \frac{1}{J} \sum_{\lambda \in \Lambda} \frac{\left\| \bfg_{\lambda}^{[t_{\lambda}]}  - \bfg_{\lambda}^{[t_{\lambda}-1]} \right\|_{2} }{\left\| \bfg_{\lambda}^{[t_{\lambda}]} \right\|_{2}} \text{,} \\
\epsilon_r^{[t]} &= \frac{1}{\sqrt{P}J} \sum_{\lambda \in \Lambda} \left\| \bfg_{\lambda}^{[t]} - \bfB_{\lambda} \boalpha^{[t]} \right\|_{2} \text{,} \\
 \epsilon_d^{[t]} &= \frac{1}{\sqrt{PJ}} \left\| \boalpha^{[t]} - \boalpha^{[t-1]} \right\|_{2} \text{.}
\end{align}
The primal residual depicts the error between the local solution and the predicted consensus value. On the other hand, the dual residual depicts the convergence of the global variable $\boalpha$.

\begin{figure}[t]
\centering
 \begin{tikzpicture}[]
  \begin{semilogyaxis}[height=5.5cm, width=8.5cm, enlarge x limits=false, xlabel= $t$-th iteration, ylabel= {$\epsilon_d^{[t]}$}, legend style={font=\tiny}, legend columns=1,
  label style={font=\small}, grid=both, ticks=both, legend entries={{$K=2, \rho=5\cdot10^3 P$}, {$K=3, \rho=5\cdot10^2 P$}, {$K=3, \rho=5\cdot10^3 P$}, {$K=3, \rho=5\cdot10^4 P$} , {$K=4, \rho=5\cdot10^3 P$}}, mark options={solid}, 
   ytick={1, 0.1, 0.01 ,0.001, 0.0001 , 0.00001, 0.000001, 0.0000001}]
    \addplot+[dashed, mark=star, draw = red, thick] table[x=iter,y=r1] {dual_residual.txt};
    \addplot+[dashed, mark=asterisk, draw = cyan, thick] table[x=iter,y=r2] {dual_residual.txt};
    \addplot+[dashed, mark=diamond, draw = blue, thick] table[x=iter,y=r3] {dual_residual.txt};
    \addplot+[dashed, mark=o, draw = olive, thick] table[x=iter,y=r4] {dual_residual.txt};
    \addplot+[dashed, mark=+ , draw = orange, thick] table[x=iter,y=r5] {dual_residual.txt};
   \end{semilogyaxis}
  \end{tikzpicture}
 \caption{Dual residual, $\epsilon_d$, as function of the iteration number $t$ of \fref{alg:gain_estimation}, for smoothing polynomial terms $K = 2, 3, 4$ and regularization term $\rho = 5\cdot10^2 P, 5\cdot10^3 P, 5\cdot10^4 P$.\label{fig:results_g_2}}
\end{figure}
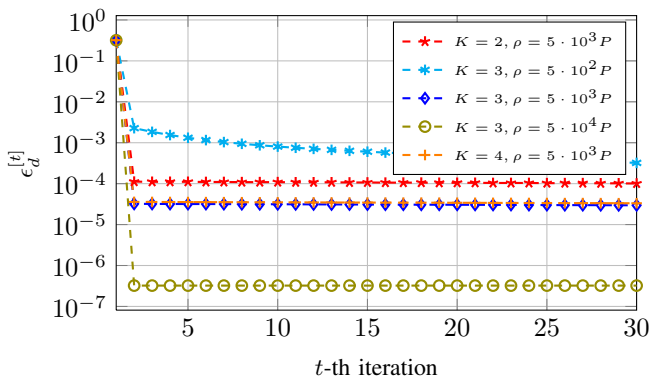

\begin{figure}[ht]
\centering
 \begin{tikzpicture}[]
   \begin{axis}[height=5.5cm, width=8.5cm, enlarge x limits=false, xlabel= Wavelength (m), ylabel= RMSE on $\bfg_{\lambda}$, legend style={font=\tiny, at={(0.45,-0.3)}, anchor = north}, legend columns=2,
   label style={font=\small},grid=both, ticks=both, legend entries={{$\bfC_{\lambda}$}, {$\bfC$}, {mono-calibration}, {$K=2, \rho=5\cdot10^3 P$}, {$K=3, \rho=5\cdot10^2 P$}, {$K=3, \rho=5\cdot10^3 P$}, {$K=3, \rho=5\cdot10^4 P$} , {$K=4, \rho=5\cdot10^3 P$}},
  xtick={7.5, 10, 12.5, 15},
    y tick label style={/pgf/number format/.cd,fixed,fixed zerofill, precision=2}, mark options={solid}]
    \addplot+[mark=none, draw = gray, thick] table[x=lambda,y=crb] {err_gain_F.txt};
    \addplot+[mark=none, draw = black, thick] table[x=lambda,y=multicrb] {err_gain_F.txt};
    \addplot+[mark=square, draw = yellow, thick] table[x=lambda,y=r0] {err_gain_F.txt};
    \addplot+[dashed,mark=star, draw = red, thick] table[x=lambda,y=r1] {err_gain_F.txt};
    \addplot+[dashed, mark=asterisk, draw = cyan, thick] table[x=lambda,y=r2] {err_gain_F.txt};
    \addplot+[dashed, mark=diamond, draw = blue, thick] table[x=lambda,y=r3] {err_gain_F.txt};
    \addplot+[dashed, mark=o, draw = olive, thick] table[x=lambda,y=r4] {err_gain_F.txt};
    \addplot+[dashed, mark=+ , draw = orange, thick] table[x=lambda,y=r5] {err_gain_F.txt};
   \end{axis}
  \end{tikzpicture}
 \caption{RMSE on $\bfg_{\lambda}$ as function of wavelength and compared to the CBRs. Parallel calibration with $K = K^{\TRUE}$ obtains the lowest error, with $\rho = 5\cdot10^3 P$ and $\rho = 5\cdot10^4 P$. The edge wavelengths have a higher error, particularly for $K = 2$ and $K = 4$, due to our choice of false interpolating polynomials.\label{fig:results_g_3}}
\end{figure}
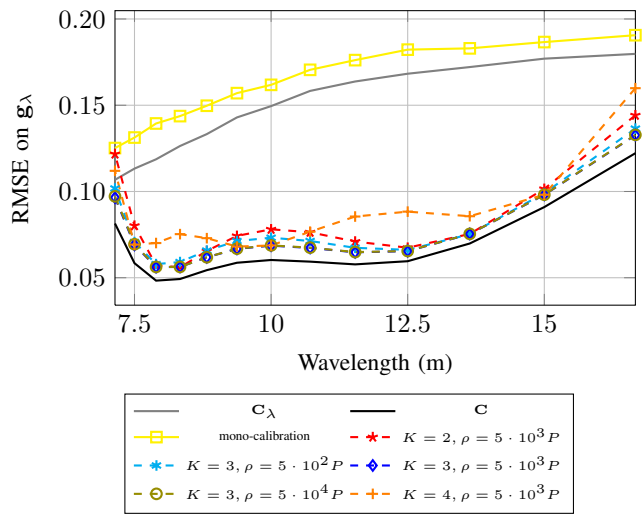

In \Fref{fig:results_g_5}, we focus on the convergence of \fref{alg:step_g}.
The $\bfg_{\lambda}$-residual decreases strongly until $t_{\lambda} \sim 10$ iterations for $t = 1$, mainly because we have a quite poor intial estimate. For $t>1$, the convergence is almost immediate ($t_{\lambda} \sim 5$ iterations). In \Fref{fig:results_g_1} and \Fref{fig:results_g_2}, we show, respectively, the primal and dual residuals, both as function of the $t$-th iteration number, when $N = 2^{14}$. We have set the regularization term $\rho = 5\cdot10^2 P, 5\cdot10^3 P, 5\cdot10^4 P$ and the smoothing polynomial order $K = 2,3,4$, with $K = 2$ underestimating the simulated polynomial order while $K = 4$ overestimating it. It is clear that as the value of $\rho$ increases, the primal and dual residuals converge faster, for  $t \sim 5$ iterations, for a sufficient value of $\rho$. Meanwhile, the primal and dual residuals differ slightly for different polynomial order $K$.

The statistical performance is then compared with mono-calibration scheme,  $\bfC_{\lambda}$ and the multi-constrained-CRB, $\bfC$. In \Fref{fig:results_g_3}, we plot the Root Mean Square Error (RMSE) for the estimates of $\bfg_{\lambda}$. The number of observations is kept to $N = 2^{14}$ and results are averaged for 500 Monte-Carlo simulations, for each chosen value of $K$ and $\rho$. We approach the multi-constrained-CRB for $K = K^{\TRUE}$ and even with both $K = 2$ and $K = 4$, we significantly improve mono-calibration. Moreover, we also have errors due to polynomial interpolation, which is clearly seen at the edge wavelengths.

\subsubsection{Results for the PCA}
we similarly analyze both convergence and performance of the proposed PCA. During the DoA estimation, we choose initially a coarse grid, with the same resolution for each coordinate of each calibrator. We apply grid refinements \cite{Mal2005} until we avoid off-grid mismatch.

\begin{figure}[ht]
\centering
 \begin{tikzpicture}[]
  \begin{semilogyaxis}[height=5.5cm, width=8.5cm, enlarge x limits=false, xlabel= $k$-th iteration, ylabel= $\epsilon_{\tbfm}^{[k]}$, legend style={font=\tiny}, legend columns=1, grid=both, ytick={1e-1, 1e-2, 1e-3, 1e-4, 1e-5},
   label style={font=\small}, ticks=both, legend entries={{$i=1$}, {$i=2$}, {$i=3$}, {$i=4$}, {$i=5$}},
   mark options={solid}]
   \addplot+[dashed, mark=star, draw = red, thick] table[x=iter,y=i1] {tbomu_residual.txt};
   \addplot+[dashed, mark=|, draw = cyan, thick] table[x=iter,y=i2] {tbomu_residual.txt};
   \addplot+[dashed, mark=diamond , draw = blue, thick] table[x=iter,y=i3] {tbomu_residual.txt};
   \addplot+[dashed, mark=o, draw = olive, thick] table[x=iter,y=i4] {tbomu_residual.txt};
   \addplot+[dashed, mark=+, draw = orange, thick] table[x=iter,y=i5] {tbomu_residual.txt};
   \end{semilogyaxis}
  \end{tikzpicture}
 \caption{$\tbfm$-residual, $\epsilon_{\tbfm}$, as function of the iteration number $k$ of \fref{alg:iht}, for different values of the $i$-th iteration of \fref{alg:proposed}\label{fig:results_doa_3}.}
\end{figure}
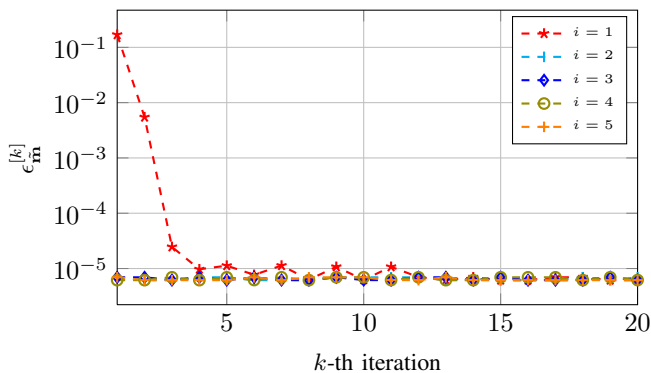

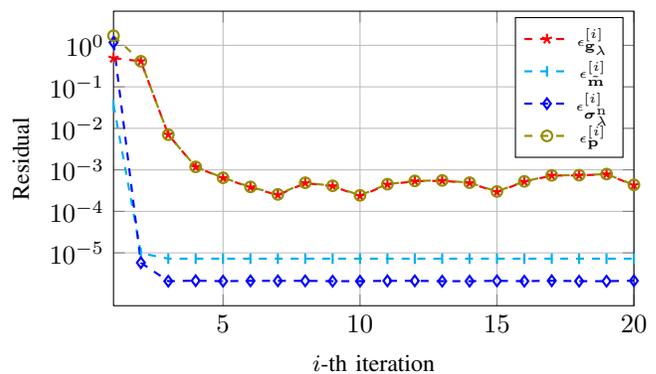
\begin{figure}[ht]
\centering
 \begin{tikzpicture}[]
  \begin{semilogyaxis}[height=5.5cm, width=8.5cm, enlarge x limits=false, xlabel= $i$-th iteration, ylabel= Residual, legend style={font=\tiny}, legend columns=1, grid=both, ytick={1, 1e-1, 1e-2, 1e-3, 1e-4, 1e-5}, label style={font=\small}, ticks=both, legend entries={{$\epsilon_{\bfg_{\lambda}}^{[i]}$}, {$\epsilon_{\tbfm}^{[i]}$}, {$\epsilon_{\bosigma^{\rmn}_{\lambda}}^{[i]}$}, {$\epsilon_{\bfp}^{[i]}$}},
   mark options={solid}]
   \addplot+[dashed, mark=star, draw = red, thick] table[x=iter,y=g] {i_residual.txt};
   \addplot+[dashed, mark=|, draw = cyan, thick] table[x=iter,y=tbomu] {i_residual.txt};
   \addplot+[dashed, mark=diamond , draw = blue, thick] table[x=iter,y=sigman] {i_residual.txt};
   \addplot+[dashed, mark=o, draw = olive, thick] table[x=iter,y=p] {i_residual.txt};
   \end{semilogyaxis}
  \end{tikzpicture}
 \caption{Variation of the residuals as function of the iteration number $i$ of \fref{alg:proposed}\label{fig:results_doa_4}.}
\end{figure}

Firstly, we concentrate on the convergence of \fref{alg:proposed} and \fref{alg:iht}, respectively. For this purpose, we define the $\tbfm$, $\bosigma_{\lambda}^{\rmn}$ and $\bfp$-residuals, respectively, by
\begin{align}
\epsilon^{[k]}_{\tbfm} &= \frac{1}{J}  \sum_{\lambda \in \Lambda} \Bigg( \frac{\left\| \bfm^{[k]}_{\lambda} - \bfm^{[k-1]}_{\lambda} \right\|_2}{\left\| \bfm^{[k]}_{\lambda} \right\|_2} + \nonumber \\ 
&\quad \frac{1}{Q} \sum_{q=1}^{Q} \left\| \bfd^{[k]}_{\lambda, q} - \bfd^{[k-1]}_{\lambda, q} \right\|_2 \Bigg)  \text{,}\\
 \epsilon^{[i]}_{\bosigma^{\rmn}_{\lambda}} &= \frac{1}{J} \sum_{\lambda \in \Lambda} \frac{ \left\| \bosigma_{\lambda}^{\rmn[i]} - \bosigma_{\lambda}^{\rmn[i-1]} \right\|_2}{ \left\| \bosigma_{\lambda}^{\rmn[i]}\right\|_2} \text{,}\\
  \epsilon^{[i]}_{\bfp} &= \epsilon_{\bfg_{\lambda}}^{[i]} + \epsilon^{[i]}_{\tbfm} + \epsilon^{[i]}_{\bosigma^{\rmn}_{\lambda}} \text{.}
\end{align}
In \Fref{fig:results_doa_3}, the $\tbfm$-residual for \fref{alg:iht} decreases during the first iterations ($k \sim 5$) and stops due to alternating between close directions on the grid. In \Fref{fig:results_doa_4}, the previous residuals and $\bfp$-residual decline more slowly and we have to wait $i \sim 10$ iterations to assure a correct convergence.

\begin{figure}[ht]
\centering
    \begin{tikzpicture}[]
   \begin{loglogaxis}[
    height=5.5cm, width=8.5cm, enlarge x limits=false, xlabel=$N$,ylabel= Root Mean Square Error, 
     legend entries={{$\bfC$-$\bfD_{\lambda}$}, {$\bfC$-$\bfm_{\lambda}$}, {RMSE-$\bfD_{\lambda}$}, {RMSE-$\bfm_{\lambda}$}},
    label style={font=\small},grid=both, ticks=both, legend style={font=\tiny}, legend columns=1,
   mark options={solid}]
   \addplot+[mark=star, draw = black, thick] table[x=n,y=d] {multicrb.txt};
   \addplot+[mark=o, draw = black, thick] table[x=n,y=mu] {multicrb.txt};
   \addplot+[dashed, mark=star, draw = red, thick] table[x=n,y=d] {rmse.txt};
   \addplot+[dashed, mark=o, draw = blue, thick] table[x=n,y=mu] {rmse.txt};
   \end{loglogaxis}
  \end{tikzpicture}
  \caption{RMSE on the directions of the calibrators and their associated directional gains as function of number of sample $N$, and compared to their corresponding multi-constrained-CRB. \label{fig:results_doa_2}}
\end{figure}
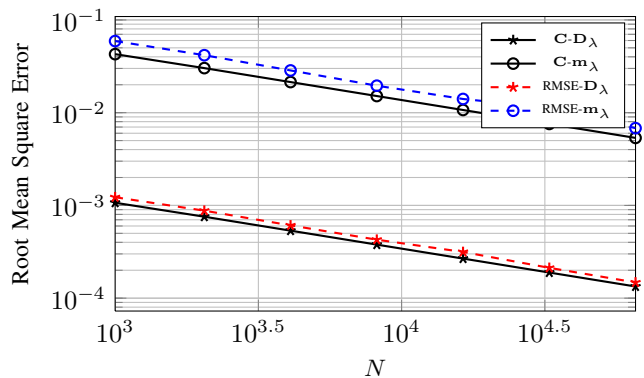

\begin{figure}[ht]
\centering
  \begin{tikzpicture}[]
   \begin{loglogaxis}[
    height=5.5cm, width=8.5cm, enlarge x limits=false, xlabel=$N$,ylabel= Root Mean Square Error, ytick={1e-0, 1e-1, 1e-2}, minor ytick={0,0.01,...,0.1,0.1,0.2,...,1},
    legend entries={{$\bfC$-$\bfg_{\lambda}$}, {$\bfC$-$\bosigma_{\lambda}^{\rmn}$(relative)}, {RMSE-$\bfg_{\lambda}$},   {RMSE-$\bosigma_{\lambda}^{\rmn}$(relative)}}, 
    label style={font=\small},grid=both, ticks=both, legend style={font=\tiny}, legend columns=2, ymax=1,
   mark options={solid}]
   \addplot+[mark=star, draw = black, thick] table[x=n,y=g] {multicrb.txt};
   \addplot+[mark=diamond, draw = black, thick] table[x=n,y=sigma_n] {multicrb.txt};
   \addplot+[dashed, mark=star, draw = red, thick] table[x=n,y=g] {rmse.txt};
   \addplot+[dashed, mark=diamond, draw = green, thick] table[x=n,y=sigma_n] {rmse.txt};
   \end{loglogaxis}
  \end{tikzpicture}
  \caption{RMSE on the undirectional gains and antenna noise powers as function of number of sample $N$, and compared to their corresponding multi-constrained-CRB. \label{fig:results_doa_1}}
\end{figure}
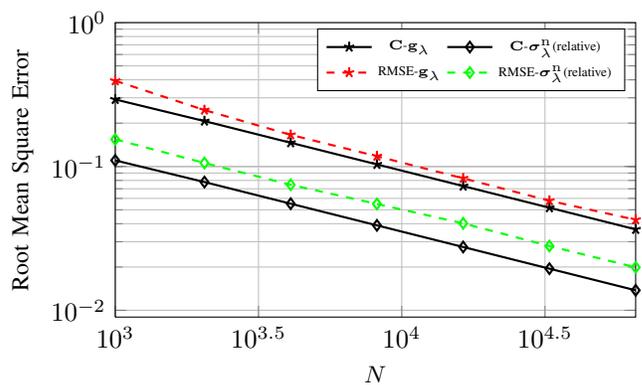

In order to investigate the statistical performances, we perform 200 Monte-Carlo runs for different sample sizes $N$, after setting $K=K^{\TRUE}, \rho = 5\cdot10^3 P$. We plot the RMSE on the different parameters in \Fref{fig:results_doa_2} and \Fref{fig:results_doa_1}, as function of the number of samples $N$ and compared to their corresponding multi-constrained-CRB. As expected, the method approaches the multi-constrained-CRB. This clearly show the good robustness of the method in low SNR scenario with a presence of non-calibrator sources.


\section{Conclusion}
\label{sec:conclusion}

In this paper, we proposed a novel iterative scheme for parallel calibration of next generation radio interferometers, where different shifts affect the apparent directions of the calibration sources and parameter values vary across wavelength. The proposed algorithm, named Parallel Calibration Algorithm, iteratively estimates the complex undirectional antenna gains and their noise powers, whereas, it jointly estimates the directions of the calibrators and their associated direction gain. These two main steps are, respectively, based on Alternating Direction of Multiple Multipliers and Distributed Iterative Hard Thresholding procedures. This leads to a statistically efficient, computationally reasonable and robust scheme as shown by numerical simulations and compared to the newly derived constrained Cramér-Rao bound.
In complement, the fusion center could be eliminated in a scheme in which agents only exchange data with their neighbours. Additionally, when the data volume per compute agent is too large, a multiplexing scheme in which each agent alternates the data used in calibration, and yet calibrates the full dataset, could be investigated.

\bibliographystyle{IEEEtran}
\bibliography{biblio}

\end{document}